\RequirePackage{fix-cm} 
\documentclass[a4paper, twoside, reqno, dvips, 12pt]{amsart}
\usepackage{fixltx2e}   

\usepackage[latin1]{inputenc}
\usepackage[T1]{fontenc}

\usepackage{eucal}
\usepackage{esint}
\usepackage{dsfont}
\usepackage{xspace}
\usepackage{amsgen}
\usepackage{amsthm}
\usepackage{amssymb}
\usepackage{amsmath}
\usepackage{upgreek}
\usepackage{amsfonts}
\usepackage{MnSymbol}
\usepackage{mathrsfs}
\usepackage{mathtools}
\usepackage[nice]{nicefrac}

\usepackage{a4wide}

\usepackage[dvipsnames, table]{xcolor}

\definecolor{oneblue}{rgb}{0.0, 0.0, 0.85}
\definecolor{bluepigment}{rgb}{0.2, 0.2, 0.6}
\definecolor{darkgrey}{rgb}{0.273, 0.281, 0.30}
\definecolor{Lightgray}{rgb}{0.89, 0.89, 0.89}
\definecolor{Lightblue}{RGB}{214, 214, 214}
\definecolor{bckg}{RGB}{20.8, 20.8, 20.8} 
\definecolor{charcoal}{rgb}{0.21, 0.27, 0.31}
\definecolor{denimblue}{rgb}{0.08, 0.38, 0.74}
\definecolor{darkelectricblue}{rgb}{0.33, 0.41, 0.47}

\usepackage{psfrag}
\usepackage{graphicx}
\usepackage{subfigure}
\usepackage{morefloats}
\usepackage{indentfirst}

\usepackage{acronym}
\usepackage{microtype}
\usepackage[labelsep=period,%
            labelfont={bf,sf,color=bluepigment},%
            justification=raggedright]{caption}

\usepackage[usenames, dvipsnames, pdf]{pstricks}
\usepackage{epsfig}
\usepackage{pst-grad} 
\usepackage{pst-plot} 

\usepackage[colorlinks,
            backref=page,
            urlcolor=oneblue,
            linkcolor=oneblue,
            citecolor=NavyBlue,
            hypertexnames=true,
            bookmarksopen=false,
            pdfpagemode=UseNone,
            pagebackref=true]{hyperref}

\usepackage[sort&compress, comma, square, numbers]{natbib}

\usepackage[explicit]{titlesec}

\titleformat{\section}
  {\color{NavyBlue}\Large\sffamily\bfseries}
  {}
  {0em}
  {\colorbox{bckg!5}{\parbox{\dimexpr\linewidth-2\fboxsep\relax}{\centering\thesection. #1}}}
  [\vspace*{0.33em}]

\titleformat{name=\section,numberless}
  {\color{NavyBlue}\Large\sffamily\bfseries}
  {}
  {0.0em}
  {\colorbox{bckg!10}{\parbox{\dimexpr\linewidth-2\fboxsep\relax}{\centering#1}}}
  [\vspace*{0.33em}]

\titleformat{\subsection}
  {\color{NavyBlue}\large\sffamily\bfseries}
  {}
  {0.0em}
  {\colorbox{bckg!5}{\parbox{\dimexpr\linewidth-2\fboxsep\relax}{\centering\thesubsection. #1}}}
  [\vspace*{0.33em}]

\titleformat{name=\subsection,numberless}
  {\color{NavyBlue}\Large\sffamily\bfseries}
  {}
  {0em}
  {\colorbox{bckg!10}{\parbox{\dimexpr\linewidth-2\fboxsep\relax}{\centering#1}}}
  [\vspace*{0.33em}]

\titleformat{\subsubsection}
  {\color{bluepigment}\sffamily\normalsize\bfseries}
  {\thesubsubsection}
  {0.5em}
  {#1}
  [\vspace*{0.33em}]

\titleformat{\paragraph}[runin]
  {\color{bluepigment}\sffamily\small\bfseries}
  {}
  {0em}
  {#1}

\titlespacing{\section}{1.0em}{1.5em plus 2pt minus 2pt}{1.0em plus 2pt minus 2pt}[0em]
\titlespacing{\subsection}{1.0em}{1.5em plus 2pt minus 2pt}{1.0em}[0em]
\titlespacing{\subsubsection}{1.0em}{1.5em plus 2pt minus 2pt}{1.0em plus 2pt minus 2pt}[0em]

\usepackage{titletoc}

\contentsmargin{0.5em}
\newlength{\tocsep} 
\titlecontents{section}[\tocsep]
  {\addvspace{10pt}\bfseries\sffamily}
  {\contentslabel[\thecontentslabel]{\tocsep}}
  {}
  {\ \titlerule*[0.75pc]{.}\ \thecontentspage}
  []
\titlecontents{subsection}[\tocsep]
  {\addvspace{8pt}\sffamily}
  {\contentslabel[\thecontentslabel]{\tocsep}}
  {}
  {\ \titlerule*[0.5pc]{.}\ \thecontentspage}
  []
\titlecontents*{subsubsection}[\tocsep]
  {\addvspace{2pt}\footnotesize\sffamily}
  {}
  {}
  {\ \titlerule*[0.35pc]{.}\ \thecontentspage}
  [\\*]

\makeatletter
\def\@setauthors{%
  \begingroup
  \def\thanks{\protect\thanks@warning}%
  \trivlist
  \centering\footnotesize \@topsep30\p@\relax
  \advance\@topsep by -\baselineskip
  \item\relax
  \author@andify\authors
  \def\\{\protect\linebreak}%
  \textsc{\normalsize\textcolor{darkelectricblue}{\authors}}%
  \ifx\@empty\contribs
  \else
    ,\penalty-3 \space \@setcontribs
    \@closetoccontribs
  \fi
  \endtrivlist
  \endgroup
}
\def\@settitle{\begin{center}%
  \baselineskip14\p@\relax
    \bfseries
    \textsc{\Large\textcolor{charcoal}{\@title}}
  \end{center}%
}
\makeatother

\usepackage{enumitem}
\setlist[description]{%
  topsep=30pt,               
  itemsep=5pt,               
  font={\bfseries\sffamily\color{bluepigment}}, 
}

\usepackage{fancyhdr}
\usepackage{lastpage}

\newcommand*\Title{\textcolor{bluepigment}{Modified Shallow Water Equations}}
\newcommand*\Authors{\textcolor{bluepigment}{D.~Dutykh \& D.~Clamond}}
\newcommand*{\plogo}{\textcolor{gray}{{\texttt{arXiv.org} / \textsc{hal}}}} 

\numberwithin{equation}{section}

\newtheorem{remark}{Remark}

\newcommand{\um}{\bar{\/u}}
\newcommand{\wb}{\bar{\/w}}
\newcommand{\A}{\mathbb{A}}
\newcommand{\R}{\mathds{R}}
\newcommand{\Z}{\mathds{Z}}
\newcommand{\J}{\mathbb{J}}
\newcommand{\vb}{\check{v}}
\newcommand{\C}{\mathcal{C}}
\newcommand{\F}{\mathcal{F}}
\newcommand{\ud}{\mathrm{d}}
\newcommand{\dt}{\partial_t}
\newcommand{\dy}{\partial_y}
\renewcommand{\beta}{\upbeta}
\newcommand{\Fr}{\mathrm{Fr}}
\renewcommand{\H}{\mathcal{H}}
\renewcommand{\L}{\mathcal{L}}
\renewcommand{\S}{\mathcal{S}}
\renewcommand{\O}{\mathcal{O}}

\newcommand{\nus}{\tilde{\nu}}

\newcommand{\nub}{\check{\nu}}
\newcommand{\phim}{\bar{\phi}}
\renewcommand{\alpha}{\upalpha}
\newcommand{\x}{\boldsymbol{x}}                     
\newcommand{\phis}{\tilde{\phi}}
\newcommand{\phib}{\check{\phi}}
\renewcommand{\u}{\boldsymbol{u}}                   
\newcommand{\sign}{\mathrm{sign}}
\newcommand{\diag}{\mathrm{diag}}
\newcommand{\sech}{\,\mathrm{sech}}
\newcommand{\vmu}{\boldsymbol{\mu}}
\newcommand{\od}[2]{\frac{\mathrm{d}#1}{\mathrm{d}#2}}
\renewcommand{\epsilon}{\varepsilon}
\newcommand{\omb}{\boldsymbol{\omega}}
\newcommand{\scal}{\boldsymbol{\cdot}}              
\newcommand{\nab}{\boldsymbol{\nabla}}              
\newcommand{\vum}{\bar{\boldsymbol{\/u}}}
\newcommand{\half}{{\textstyle{1\over2}}}

\newcommand{\frth}{{\textstyle{3\over4}}}

\newcommand{\vmus}{\tilde{\boldsymbol \mu}}
\newcommand{\vmub}{\check{\boldsymbol \mu}}
\newcommand{\onethird}{{\textstyle{1\over3}}}
\newcommand{\minmod}{\mathop{\mathrm{minmod}}}
\newcommand{\twothirds}{{\textstyle{2\over3}}}
\newcommand{\pd}[2]{\frac{\partial #1}{\partial #2}}

\newcommand{\ie}{\emph{i.e.}\/ }
\newcommand{\eg}{\emph{e.g.}\/ }

\usepackage{acronym}
\acrodef{msv}[mSV]{modified Saint-Venant}

\begin{document}

\title[\Title]{Modified Shallow Water Equations for significantly varying seabeds}

\author[D.~Dutykh]{Denys~Dutykh$^*$}
\address{Universit\'e Savoie Mont Blanc, LAMA, UMR 5127 CNRS, Campus Scientifique, 73376 Le Bourget-du-Lac Cedex, France}
\email{Denys.Dutykh@univ-savoie.fr}
\urladdr{http://www.denys-dutykh.com/}
\thanks{$^*$ Corresponding author}

\author[D.~Clamond]{Didier~Clamond}
\address{Universit\'e de Nice -- Sophia Antipolis, Laboratoire J.~A. Dieudonn\'e, Parc Valrose, 06108 Nice cedex 2, France}
\email{diderc@unice.fr}
\urladdr{http://math.unice.fr/~didierc/}


\begin{titlepage}
\thispagestyle{empty} 
\noindent
{\Large Denys \textsc{Dutykh}}\\
{\it\textcolor{gray}{CNRS, Universit\'e Savoie Mont Blanc, France}}
\\[0.02\textheight]
{\Large Didier \textsc{Clamond}}\\
{\it\textcolor{gray}{Universit\'e de Nice -- Sophia Antipolis, France}}\\[0.16\textheight]

\vspace*{0.99cm}

\colorbox{Lightblue}{
  \parbox[t]{1.0\textwidth}{
    \centering\huge\sc
    \vspace*{0.79cm}
    
    \textcolor{bluepigment}{Modified Shallow Water Equations for significantly varying seabeds}

    \vspace*{0.79cm}
  }
}

\vfill 

\raggedleft     
{\large \plogo} 
\end{titlepage}


\newpage
\thispagestyle{empty} 
\par\vspace*{\fill}   
\begin{flushright} 
{\textcolor{denimblue}{\textsc{Last modified:}} \today}
\end{flushright}


\newpage
\maketitle
\thispagestyle{empty}


\begin{abstract}
In the present study, we propose a modified version of the Nonlinear Shallow Water Equations (Saint-Venant or NSWE) for irrotational surface waves in the case when the bottom undergoes some significant variations in space and time. The model is derived from a variational principle by choosing an appropriate shallow water ansatz and imposing some constraints. Our derivation procedure does not explicitly involve any small parameter and is straightforward. The novel system is a non-dispersive non-hydrostatic extension of the classical Saint-Venant equations. A key feature of the new model is that, like the classical NSWE, it is hyperbolic and thus similar numerical methods can be used. We also propose a finite volume discretisation of the obtained hyperbolic system. Several test-cases are presented to highlight the added value of the new model. Some implications to tsunami wave modelling are also discussed.

\bigskip
\noindent \textbf{\keywordsname:} Shallow water; Saint-Venant equations; finite volumes; UNO scheme \\


\end{abstract}

\newpage
\tableofcontents
\thispagestyle{empty}


\newpage
\section{Introduction}

The celebrated classical nonlinear shallow water equations were derived in 1871 by A.J.C. de~\textsc{Saint-Venant} \cite{SV1871}. Currently these equations are widely used in practice and one can find thousands of publications devoted to the applications, validations and numerical solutions of these equations \cite{Dutykh2010c, Dutykh2009a, Medeiros2013}.

The interaction of surface waves with mild or tough bottoms has always attracted the particular attention of researchers \cite{Artiles2004, Chazel2007, Dutykh2011c, Nachbin2003}. There are however few studies which attempt to include the bottom curvature effect into the classical Saint-Venant \cite{SV1871, Stoker1957} or Savage--Hutter\footnote{The Savage--Hutter equations are usually posed on inclined planes and they are used to model various gravity driven currents, such as snow avalanches \cite{Ancey2006}.} \cite{Gray1998,Savage1989} equations. One of the first studies in this direction is perhaps due to \textsc{Dressler} \cite{Dressler1978}. Much later, this research was pursued almost in the same time by \textsc{Berger, Keller, Bouchut} and their collaborators \cite{Berger1998, Bouchut2003, Keller2003}. We note that all these authors used some variants of the asymptotic expansion method. Recently, the model proposed by \textsc{Dressler} was validated in laboratory experiments \cite{Dewals2006}. The present study is a further attempt to improve the classical Saint-Venant equations by including a better representation of the bottom shape. \textsc{Dressler}'s model includes the bottom curvature effects, which require the computation of bottom's profile second order derivatives. For irregular shapes it can be problematic. Consequently, we try below to propose a model which requires only first spatial derivatives of the bathymetry.

The Saint-Venant equations are derived under the assumption of a hydrostatic pressure field, resulting in a non-dispersive system of equations. Many non-hydrostatic improved models have long been proposed, see \cite{Beji1997, Nadaoka1997, Madsen1999, Wu2001a} for reviews. These Boussinesq-like and/or mild-slope \cite{Beji1997, Nadaoka1997} equations are dispersive (\ie, the wave speed depends on the wavelength) and involve (at least) third-order derivatives. Although these models capture more physical effects than the classical Saint-Venant shallow water equations, they have several drawbacks. First, the dispersive effects are often negligible for very long waves such as tsunamis and tide waves. Second, the higher-order derivatives introduce stiffness into the equations and thus their numerical resolution is significantly more involved and costly than for the Saint-Venant equations. Third, the Boussinesq-like equations are not hyperbolic and, unlike the Saint-Venant equations, the method of characteristics cannot be employed (unless the operators are splitted, \eg \cite{Bonneton2011}). Therefore, it is not surprising that various dispersive shallow water models are not systematically used in coastal modelling.

In presence of a varying bathymetry, the shallow water equations are derived under the assumption that the bottom variations are very weak. However, even for very long surface waves, significant variations of the bathymetry can play an important role in the wave propagation. These bottom slope effects can be even more important when the wave travels over many oscillations of the seabed, due to the accumulation of bottom slope influences. Therefore, even for a shallow water long waves  model, it is important to take properly into account the significant bottom variations \cite{Chazel2007}. In this article, we present a modification of the Saint-Venant equations in presence of a seabed of significant variations. This model is derived from a variational principle, which is a powerful method to derive approximations that cannot be obtained from more classical asymptotic expansions.

In the theory of water waves, variational principles are generally used together with small parameter expansions. Doing so, the approximations derived are identical to the one obtained from asymptotic expansions directly used into the equations. Thus, the only advantage of a variational method is elegance and simplified derivations. However, variational methods are much more powerful than that and approximations can also be obtained without relying on asymptotic expansions. This is specially useful when no obvious small parameter can be identified in the problem at hands.

Indeed, variational methods have been more popular in Physics \cite{Basdevant2007}, especially in Quantum Mechanics \cite{Sakurai1993} than in Fluid Mechanics where the majority of approximate model derivations use the perturbation-type techniques. The main reason for this discrepancy comes probably from the fact that in most problems of Quantum Mechanics a small parameter cannot be simply identified (roughly speaking everything scales with the Planck constant $\hbar$). Consequently, physicists had to develop alternative methods based on the guess of the solution structure, translated into the mathematical language as the so-called solution's \emph{ansatz}. For example, a particularly good guess of the ansatz was made by R.~\textsc{Laughlin} \cite{Laughlin1983} for the quantum Hall effect, which was distinguished 15 years later by the Nobel Prize in Physics in 1998. 

Here, we adopt the same philosophy applying it to the long water waves propagating over a seabed with significant variations. Namely, the shallow water ansatz from \cite{Clamond2009} is additionally constrained to respect the bathymetry variations in space and time. Then, applying the variational principle, we arrive naturally to some \acf{msv} equations. These mSV equations, like the classical Saint-Venant equations, are hyperbolic and can be solved with similar techniques, that is an interesting feature in the prospect of integration/modification of existing operational codes. The derivation of \acs{msv} equations presented below were communicated by the same authors in a short note announcing the main results \cite{Dutykh2011b}. In the present study, we investigate deeper the properties of the proposed \acs{msv} system along with its solutions through analytical and numerical methods. We specially focus on some predictions of interest for ocean modelling, in particular the fact that the waves are slowed down by the seabed slope.

This article is organised as follows. After some introductory remarks, the paper begins with the derivation and discussion of some properties of the \acf{msv} equations in Section~\ref{sec:model}. Then, we investigate the hyperbolic structure and present a finite volume scheme in Section~\ref{sec:fv}. Several numerical results are shown in Section~\ref{sec:numres}. Finally, some main conclusions and perspectives are outlined in the last Section~\ref{sec:concl}.


\section{Mathematical model}\label{sec:model}

\begin{figure}
	\centering
	\scalebox{0.9} 
	{
	\begin{pspicture}(0,-3.545)(16.039062,3.454)
	\definecolor{color32}{rgb}{0.12156862745098039,0.08627450980392157,0.8352941176470589}
	\definecolor{color96}{rgb}{0.3254901960784314,0.1607843137254902,0.1607843137254902}
	\definecolor{color145}{rgb}{0.054901960784313725,0.01568627450980392,0.01568627450980392}
	\definecolor{color145b}{rgb}{0.6549019607843137,0.44313725490196076,0.44313725490196076}
	\definecolor{color168}{rgb}{0.09411764705882353,0.0392156862745098,0.0392156862745098}
	\definecolor{color168b}{rgb}{0.06274509803921569,0.023529411764705882,0.023529411764705882}
	\definecolor{color423}{rgb}{0.2549019607843137,0.27058823529411763,0.9372549019607843}
	\definecolor{color558}{rgb}{0.22745098039215686,0.11372549019607843,0.11372549019607843}
	\psline[linewidth=0.028cm,linestyle=dashed,arrowsize=0.1cm 2.0,arrowlength=1.4,
	arrowinset=0.4]{->}(0.0,1.68)(16.0,1.68)
	\psline[linewidth=0.028cm,linestyle=dashed,arrowsize=0.1cm 2.0,arrowlength=1.4,
	arrowinset=0.4]{<-}(8.0,2.9)(8.0,-3.1)
	\pscustom[linewidth=0.044,linecolor=color32]
	{
	\newpath
	\moveto(0.96,1.68)
	\lineto(1.2286792,1.68)
	\curveto(1.3630188,1.68)(1.732453,1.8)(1.9675473,1.92)
	\curveto(2.2026415,2.04)(2.67283,2.04)(2.9079244,1.92)
	\curveto(3.1430187,1.8)(3.747547,1.52)(4.116981,1.36)
	\curveto(4.486415,1.2)(5.359623,1.2)(5.863396,1.36)
	\curveto(6.36717,1.52)(7.072453,1.8)(7.2739625,1.92)
	\curveto(7.475472,2.04)(7.8784904,2.12)(8.08,2.08)
	\curveto(8.281509,2.04)(8.718113,1.84)(8.953207,1.68)
	\curveto(9.188302,1.52)(9.591321,1.32)(9.759245,1.28)
	\curveto(9.927169,1.24)(10.263018,1.32)(10.430943,1.44)
	\curveto(10.598867,1.56)(10.901132,1.76)(11.035471,1.84)
	\curveto(11.169811,1.92)(11.472075,2.0)(11.64,2.0)
	\curveto(11.807925,2.0)(12.177359,2.0)(12.378868,2.0)
	\curveto(12.580379,2.0)(12.949812,1.96)(13.117737,1.92)
	\curveto(13.285662,1.88)(13.621509,1.8)(13.7894335,1.76)
	\curveto(13.957358,1.72)(14.293208,1.68)(14.461133,1.68)
	\curveto(14.629058,1.68)(14.897737,1.68)(15.2,1.68)
	}
	\psline[linewidth=0.05cm,linecolor=color96](0.32,-2.64)(1.6281435,-2.64)
	\psline[linewidth=0.05cm,linecolor=color96](14.055508,-2.64)(15.494465,-2.64)
	\pscustom[linewidth=0.05,linecolor=color96]
	{
	\newpath
	\moveto(1.6281437,-2.64)
	\lineto(1.8897723,-2.56)
	\curveto(2.0205867,-2.52)(2.2495117,-2.4)(2.3476226,-2.32)
	\curveto(2.4457333,-2.24)(2.609251,-2.28)(2.6746583,-2.4)
	\curveto(2.7400653,-2.52)(2.8708801,-2.8)(2.9362872,-2.96)
	\curveto(3.0016944,-3.12)(3.2306194,-3.24)(3.3941376,-3.2)
	\curveto(3.5576556,-3.16)(3.7865806,-2.96)(3.8519876,-2.8)
	\curveto(3.9173946,-2.64)(4.0809126,-2.36)(4.1790233,-2.24)
	\curveto(4.2771344,-2.12)(4.5060596,-2.12)(4.6368737,-2.24)
	\curveto(4.767688,-2.36)(4.8985023,-2.68)(4.8985023,-2.88)
	\curveto(4.8985023,-3.08)(4.996613,-3.36)(5.094724,-3.44)
	\curveto(5.1928344,-3.52)(5.42176,-3.44)(5.5525746,-3.28)
	\curveto(5.6833887,-3.12)(5.912314,-2.76)(6.0104246,-2.56)
	\curveto(6.1085353,-2.36)(6.337461,-2.2)(6.468275,-2.24)
	\curveto(6.599089,-2.28)(6.7953105,-2.48)(6.860718,-2.64)
	\curveto(6.926125,-2.8)(7.1550503,-3.0)(7.318568,-3.04)
	\curveto(7.482086,-3.08)(7.711011,-3.0)(7.7764187,-2.88)
	\curveto(7.8418255,-2.76)(8.038047,-2.52)(8.168861,-2.4)
	\curveto(8.299675,-2.28)(8.528601,-2.28)(8.626712,-2.4)
	\curveto(8.724822,-2.52)(8.88834,-2.8)(8.953748,-2.96)
	\curveto(9.019155,-3.12)(9.24808,-3.2)(9.411598,-3.12)
	\curveto(9.575116,-3.04)(9.804041,-2.8)(9.869449,-2.64)
	\curveto(9.934856,-2.48)(10.098374,-2.24)(10.196485,-2.16)
	\curveto(10.294595,-2.08)(10.490815,-2.12)(10.588927,-2.24)
	\curveto(10.687038,-2.36)(10.817853,-2.64)(10.850556,-2.8)
	\curveto(10.883259,-2.96)(11.014074,-3.12)(11.1121855,-3.12)
	\curveto(11.210296,-3.12)(11.373814,-2.96)(11.43922,-2.8)
	\curveto(11.504626,-2.64)(11.668144,-2.36)(11.766256,-2.24)
	\curveto(11.8643675,-2.12)(12.027885,-2.04)(12.093292,-2.08)
	\curveto(12.158699,-2.12)(12.289515,-2.28)(12.35492,-2.4)
	\curveto(12.420327,-2.52)(12.583845,-2.68)(12.681957,-2.72)
	\curveto(12.780068,-2.76)(13.008993,-2.84)(13.139807,-2.88)
	\curveto(13.270621,-2.92)(13.499546,-2.88)(13.597656,-2.8)
	\curveto(13.695766,-2.72)(13.859284,-2.64)(14.055508,-2.64)
	}
	\usefont{T1}{ptm}{m}{n}
	\rput(15.664532,1.31){$\x$}
	\usefont{T1}{ptm}{m}{n}
	\rput(7.7,2.8){$y$}
	\usefont{T1}{ptm}{m}{n}
	\rput(7.534531,1.31){$O$}
	\psline[linewidth=0.03cm,linestyle=dotted,linecolor=color145,fillcolor=color145b,
	arrowsize=0.1cm 2.0,arrowlength=1.4,arrowinset=0.4]{<->}(11.2,1.6)(11.2,-3.0)
	\usefont{T1}{ptm}{m}{n}
	\rput(10.444531,-0.45){$d(\x,t)$}
	\psline[linewidth=0.03cm,linestyle=dotted,linecolor=color168,fillcolor=color168b,
	arrowsize=0.1cm 2.0,arrowlength=1.4,arrowinset=0.4]{<->}(4.48,1.2)(4.48,-2.1)
	\usefont{T1}{ptm}{m}{n}
	\rput(3.7245312,-0.45){$h(\x,t)$}
	\psline[linewidth=0.03cm,linestyle=dotted,linecolor=color168,fillcolor=color168b,
	arrowsize=0.1cm 2.0,arrowlength=1.4,arrowinset=0.4]{>-<}(2.4,1.45)(2.4,2.25)
	\usefont{T1}{ptm}{m}{n}
	\rput(3.1,2.3){$\eta(\x,t)$}
	\psline[linewidth=0.05cm,linecolor=color423,fillcolor=color168b](0.8,0.4)(1.12,0.4)
	\psline[linewidth=0.05cm,linecolor=color423,fillcolor=color168b](0.8,-1.68)(1.12,-1.68)
	\psline[linewidth=0.05cm,linecolor=color423,fillcolor=color168b](2.72,0.72)(3.04,0.72)
	\psline[linewidth=0.05cm,linecolor=color423,fillcolor=color168b](2.72,-1.52)(3.04,-1.52)
	\psline[linewidth=0.05cm,linecolor=color423,fillcolor=color168b](5.6,0.56)(5.92,0.56)
	\psline[linewidth=0.05cm,linecolor=color423,fillcolor=color168b](1.6,-0.4)(1.92,-0.4)
	\psline[linewidth=0.05cm,linecolor=color423,fillcolor=color168b](5.92,-1.68)(6.24,-1.68)
	\psline[linewidth=0.05cm,linecolor=color423,fillcolor=color168b](7.04,0.24)(7.36,0.24)
	\psline[linewidth=0.05cm,linecolor=color423,fillcolor=color168b](9.6,0.56)(9.92,0.56)
	\psline[linewidth=0.05cm,linecolor=color423,fillcolor=color168b](8.96,-1.68)(9.28,-1.68)
	\psline[linewidth=0.05cm,linecolor=color423,fillcolor=color168b](12.8,0.72)(13.12,0.72)
	\psline[linewidth=0.05cm,linecolor=color423,fillcolor=color168b](14.24,-1.68)(14.56,-1.68)
	\psline[linewidth=0.05cm,linecolor=color423,fillcolor=color168b](13.12,-0.56)(13.44,-0.56)
	\psline[linewidth=0.05cm,linecolor=color423,fillcolor=color168b](11.84,-1.52)(12.16,-1.52)
	\psline[linewidth=0.05cm,linecolor=color423,fillcolor=color168b](8.8,-0.4)(9.12,-0.4)
	\psline[linewidth=0.03cm,linecolor=color558,fillcolor=color168b](0.8,-2.64)(0.48,-2.96)
	\psline[linewidth=0.03cm,linecolor=color558,fillcolor=color168b](1.28,-2.64)(0.96,-2.96)
	\psline[linewidth=0.03cm,linecolor=color558,fillcolor=color168b](1.76,-2.64)(1.44,-2.96)
	\psline[linewidth=0.03cm,linecolor=color558,fillcolor=color168b](14.24,-2.64)(13.92,-2.96)
	\psline[linewidth=0.03cm,linecolor=color558,fillcolor=color168b](14.72,-2.64)(14.4,-2.96)
	\psline[linewidth=0.03cm,linecolor=color558,fillcolor=color168b](15.2,-2.64)(14.88,-2.96)
	\end{pspicture}}
\caption{\small\em Definition sketch of the fluid domain.}
\label{fig:sketch}
\end{figure}
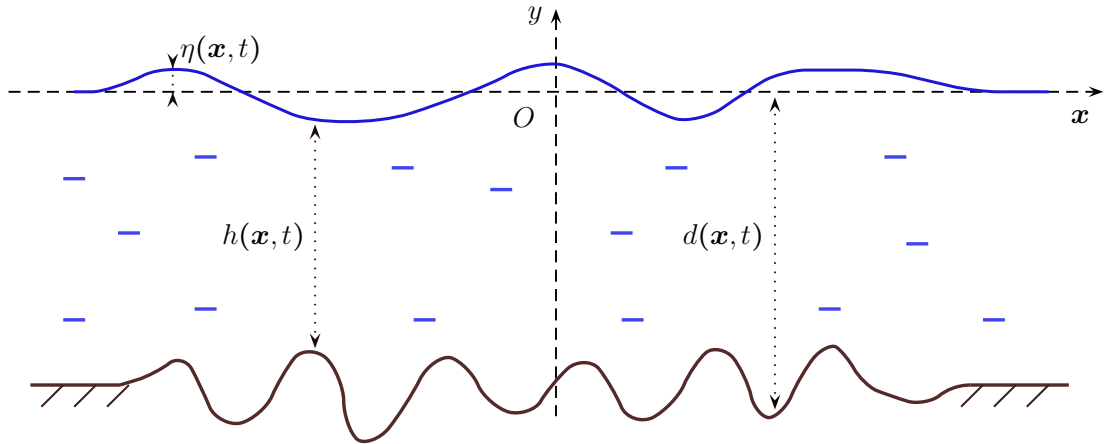

Consider an ideal incompressible fluid of constant density $\rho$. The horizontal independent variables are denoted by $\x=(x_1,x_2)$ and the upward vertical one by $y$. The origin of the Cartesian coordinate system is chosen such that the surface $y=0$ corresponds to the still water level. The fluid is bounded below by the bottom at $y = -d(\x,t)$ and above by the free surface at $y=\eta(\x,t)$. Usually, we assume that the total depth $h(\x, t)\equiv d(\x,t) + \eta(\x,t)$ remains positive $h (\x,t) \geqslant h_0 > 0$ at all times $t\in[0,T]$. The sketch of the physical domain $\Omega\times [-d,\eta]$, $\Omega\subseteq\R^2$ is shown in Figure~\ref{fig:sketch}.

Traditionally in water wave modeling the assumption of flow irrotationality is also adopted. Under these constitutive hypotheses, the governing equations of the classical water wave problem are \cite{Stoker1958}:
\begin{align}
  \nab^2\phi\ +\ \dy^{\,2}\phi\ =&\ 0, 
  \qquad (\x, y) \in \Omega\times [-d, \eta], \label{eq:laplace} \\
  \dt\eta\ +\ (\nab\phi)\scal(\nab\eta)\ -\ \dy\phi\ =&\ 0, 
  \qquad y = \eta(\x, t), \label{eq:kinematic} \\
  \dt\phi\ +\ \half|\nab\phi|^2\ +\ \half(\dy\phi)^2\ +\ g\eta\ =&\ 0, 
  \qquad y = \eta(\x,t), \label{eq:bernoulli} \\
  \dt d\ +\ (\nab d)\scal(\nab\phi)\ +\ \dy\phi\ =&\ 0, 
  \qquad y = -d(\x,t), \label{eq:bottomkin}
\end{align}
where $\phi$ is the velocity potential (by definition, $\u = \nab\phi$ and $v = \dy\phi$), $g>0$ is the acceleration due to gravity and $\nab = (\partial_{x_1}, \partial_{x_2})$ denotes the gradient operator in horizontal Cartesian coordinates.

The assumptions of fluid incompressibility and flow irrotationality lead to the Laplace equation \eqref{eq:laplace} for the velocity potential $\phi(\x,y,t)$. The main difficulty of the water wave problem lies on the boundary conditions. Equations \eqref{eq:kinematic} and \eqref{eq:bottomkin} express the free-surface kinematic condition and bottom impermeability, respectively, while the dynamic condition \eqref{eq:bernoulli} expresses the free surface isobarity.

It is well-known that the water wave problem \eqref{eq:laplace} -- \eqref{eq:bottomkin} possesses several variational structures \cite{Broer1974, Luke1967, Petrov1964, Whitham1965, Zakharov1968}. Recently, we proposed a relaxed Lagrangian variational principle which allows much more freedom in constructing approximations compared to the classical formulations. Namely, the water wave equations can be obtained as Euler--Lagrange equations of the functional $\iiint\mathscr{L}\,\ud^2\x\,\ud\/t$ involving the Lagrangian density \cite{Clamond2009}:
\begin{align}\label{defL}
\mathscr{L}\ =&\ (\partial_t\eta+\vmus\scal\nab\eta-\nus)\,\phis\ +\ (\partial_td+\vmub\scal\nab d
+\nub)\,\phib\ -\ \half\,g\,\eta^2\  \nonumber \\ &+\ \int_{-d}^{\,\eta}\left[\,\vmu\scal\u-{\half}
\/\u^2\,+\,\nu\/v-\half\/v^2\, +\,(\nab\scal\vmu+\partial_y\nu)\,\phi\,\right]\ud\/y,
\end{align}
where `over-tildes' and `wedges' denote, respectively, quantities computed at the free surface $y = \eta(\x,t)$ and at the bottom $y = -d(\x,t)$ (we shall also denote below with `bars' the quantities averaged over the water depth); \{$\u,v,\vmu,\nu$\} being the horizontal velocity, vertical velocity and associated Lagrange multipliers, respectively. The last two additional variables \{$\vmu,\nu$\} are called the pseudo-velocities. They formally arise as Lagrange multipliers associated to the constraints $\u = \nab\phi$, $v = \phi_y$. However, once these variables are introduced, the ansatz can be chosen regardless their initial definition, \ie, it is not obligatory to choose an ansatz such that the relations $\vmu = \u = \nab\phi$, for example, are exactly satisfied. The advantage of the relaxed variational principle \eqref{defL} consists in the extra freedom for constructing approximations.


\subsection{Constrained shallow water ansatz}

In order to simplify the full water wave problem, we choose some approximate but physically relevant representations of all dependent variables. In this study, we choose a simple shallow water ansatz, which is a velocity field and velocity potential independent of the vertical coordinate $y$ such that
\begin{equation}\label{anssha}
  \phi\ \approx\ \phim(\x,t), \qquad \u\ =\ \vmu\ \approx\ \vum(\x,t), \qquad v\ =\ \nu\ \approx\ \vb (\x,t),
\end{equation}
where $\vum(\x,t)$ is the depth-averaged horizontal velocity and $\vb(\x,t)$ is the vertical velocity at the bottom. In this ansatz, we take for simplicity the pseudo-velocities to be equal to the velocity field $\u = \vmu$, $v = \nu$. However, in other situations they can differ (see \cite{Clamond2009} for more examples).

Physically, the ansatz \eqref{anssha} means that we are considering a so-called {\em columnar flow\/} \cite{Miles1985}, which is a sensible model for long waves in shallow water, as long as their amplitudes are not too large. Mathematically, the ansatz \eqref{anssha} implies that the vertical variation of the velocity field does not contribute (\ie, is negligible) to the Lagrangian \eqref{defL}. Thus, with the ansatz \eqref{anssha}, the Lagrangian density \eqref{defL} becomes
\begin{align}\label{defLsw}
  \mathscr{L}\ =&\,\left(\partial_t\/h\,+\,\vum\scal\nab h\,+\,h\nab\scal\vum\right)\phim\ -\ \half\,g\,\eta^2\ +\ \half\,h\,(\vum^2+\vb^2),
\end{align}
where we introduced the total water depth $h = \eta + d$.

Since we are considering a columnar flow model, each vertical water column can be considered as a moving rigid body. In presence of bathymetry variations, the columnar flow paradigm then yields that the fluid vertical velocity must be equal to the one at the bottom, because the bottom is impermeable. Thus, we require that the fluid particles follow the bottom profile, \ie,
\begin{equation}\label{eq:bot}
  \vb\ =\ -\partial_t\,d\ -\ \vum\scal\nab d,
\end{equation}
this identity being the bottom impermeability condition expressed with the ansatz \eqref{anssha}.

\begin{remark}
Note that for ansatz \eqref{anssha} the horizontal vorticity $\omb$ and the vertical one $\zeta$ are given by:
\begin{equation*}
  \omb = \Bigl(\partial_{x_2}\vb\,,\,-\partial_{x_1}\vb\Bigr), \qquad
  \zeta = \partial_{x_1}\um_2 - \partial_{x_2}\um_1.
\end{equation*}
Consequently the flow is not exactly irrotational in general. It will be confirmed below one more time when we establish the connection between $\vum$ and $\nab\phim$.
\end{remark}

After substitution of the relation \eqref{eq:bot} into the Lagrangian density \eqref{defLsw}, the Euler--Lagrange equations yield:
\begin{align}
  \delta\/\vum\,:\quad \boldsymbol{0}\ &=\ \vum\ -\ \nab\phim\ -\ \vb\,\nab d, \label{eqdvu} \\
  \delta\/\phim\,:\quad 0\ &=\ \partial_t\,h\ +\ \nab\scal[\,h\,\vum\,], \label{eqdphi}\\
  \delta\/\eta\,:\quad 0\ &=\ \partial_t\,\phim\ +\ g\,\eta\ +\ \vum\scal\nab\phim\ -\ \half\,(\vum^2+\vb^2). \label{eqdeta}
\end{align}
Taking the gradient of \eqref{eqdeta} and eliminating of $\phim$ from \eqref{eqdvu} gives us the system of governing equations:
\begin{align}
  \partial_t\,h\ +\ \nab\scal[\,h\,\vum\,]\ &=\ 0, \label{eq:mas}\\
  \partial_t\,[\,\vum\,-\vb\,\nab d\,]\ +\ \nab\,[\,g\,\eta\ +\ \half\,\vum^2\ +\ \half\,\vb^2\ +\ \vb\,\partial_t\/d\,] &=\ 0, \label{eq:qdm}
\end{align}
together with the auxiliary relations
\begin{align} 
  \vum\ &=\ \nab\phim\ +\ \vb\,\nab d\ =\  \nab\phim\ -\ \frac{\partial_t\/d\,
  +\, (\nab\phim)\scal(\nab d)}{1\, +\, |\nab d|^2}\,\nab d, \label{equexplicit} \\
  \vb\ &=\ -\ \partial_t\,d\ -\ \vum\scal\nab d\ =\  
  -\,\frac{\partial_t\/d\, +\, (\nab\phim)\scal(\nab d)}{1\, +\, |\nab d|^2}. \label{eqvexplicit}
\end{align}
Hereafter, every times the variables $\vum$ and $\vb$ appear in equations, it is {\em always\/} assumed that they are defined by the relations \eqref{equexplicit}--\eqref{eqvexplicit}.

\begin{remark}
The classical irrotational nonlinear shallow water or Saint-Venant equations \cite{SV1871, Stoker1957} can be recovered by substituting $\vb = 0$ into the last system:
\begin{align*}
  \partial_t\,h\ +\ \nab\scal[\,h\,\vum\,]\ &=\ 0, \\
  \partial_t\,\vum\ +\ \nab\,[\,g\,\eta\ +\ \half\,\vum^2\,]\ &=\ 0,
\end{align*}
where $\vum = \nab\phim$.
\end{remark}


\subsection{Properties of the new model}

From the governing equations \eqref{eq:mas}, \eqref{eq:qdm} one can derive an equation for the horizontal velocity $\vum$:
\begin{equation}\label{eq:qdmhor}
  \partial_t\,\vum\ +\ \half\,\nab(\vum^2)\ +\ g\,\nab\eta\ =\ \gamma\,\nab d \ +\  \vum\wedge(\nab\vb\wedge\nab d),
\end{equation}
where $\gamma$ is the vertical acceleration at the bottom defined as:
\begin{equation}\label{eq:acc}
  \gamma\ \equiv\ \frac{\ud\,\vb}{\ud t}\ =\ \partial_t\,\vb\ +\ (\vum\scal\nab)\,\vb.
\end{equation}

\begin{remark}
Note that in \eqref{eq:qdmhor} the last term on the right-hand side cancels out for two-dimensional waves (\ie, one horizontal dimension). It can be seen from the following analytical representation which degenerates to zero in one horizontal dimension:
\begin{equation*}
  \vum\wedge(\nab\vb\wedge\nab d)\ =\ (\nab\vb)\,(\vum\scal\nab d)\ -\ (\nab d)\,(\vum\scal\nab\vb).
\end{equation*}
This property has an interesting geometrical interpretation since $\vum\wedge(\nab\vb\wedge\nab d)$ is a horizontal vector orthogonal to $\vum$ and thus vanishing for two-dimensional waves.
\end{remark}

Defining the depth-averaged total (kinetic plus potential) energy density $\mathscr{E}$ together with the ansatz (\ref{anssha}), \ie,
\begin{equation}\label{defenetot}
  \mathscr{E}\ =\ \int_{-d}^\eta\left[\,\frac{\u^2\/+\/v^2}{2}\,+\,g\,y\,\right]\ud\/y\ \approx\ h\,\frac{\vum^2\,+\,\vb^2}{2}\ +\ g\,\frac{\eta^2-d^2}{2},
\end{equation}
and using \eqref{equexplicit}--\eqref{eqvexplicit}, after some algebra, one derives the energy equation
\begin{equation}\label{eq:energy}
  \partial_t\,\mathscr{E}\ +\ \nab\scal\left[\,\mathscr{E}\,\vum\,+\,\half\,g\,h^2\,\vum\,\right]\, =\ -\/(g+\gamma)\,h\,\partial_td.
\end{equation}
Obviously, the source term on the right-hand side vanishes if the bottom is fixed $d = d(x)$ or, equivalently, if $\partial_t\/d = 0$.

The mSV equations \eqref{eqdvu}--\eqref{eqdeta} possess a Hamiltonian structure with canonical variables $h$ and $\phim$, \ie, 
\begin{equation*}
  \pd{\,h}{t}\ =\ \frac{\delta\,\H}{\delta\phim}, \qquad
  \pd{\,\phim}{t}\ =\ -\frac{\delta\,\H}{\delta h},
\end{equation*}
where the Hamiltonian $\H$ is defined as
\begin{equation}\label{eq:Hamilt}
  2\,\H\ =\ \int\left\{g\/(h-d)^2\,-\,g\/d^2\, +\,h\/|\nab\phim|^2\, -\,\frac{h\,[\,\partial_td\,+\,(\nab\phim)\scal(\nab d)\,]^2}{1\,+\,|\/\nab d\/|^2} \right\}\,\ud^2\x.
\end{equation}
One can easily check, after computing the variations, that the Hamiltonian \eqref{eq:Hamilt} yields
\begin{align}
  \partial_t\,h\ =&\ -\nab\scal\left[\,h\/\nab\phim\ -\ \frac{\partial_t\/d\,+\,(\nab\phim)\scal(\nab d)}{1+|\nab d|^2}\,h\,\nab d\,\right], \\
  \partial_t\,\phim\ =&\ -\/g\,(h-d)\ -\ \frac{|\/\nab\phim\/|^2}{2}\ +\ \frac{[\,\partial_t\/d\, +\, (\nab\phim)\scal(\nab d)\,]^2}{2\,+\,2\,|\/\nab d\/|^2},
\end{align}
which are equivalent to the system \eqref{eqdphi}--\eqref{eqdeta} after introduction of the auxiliary variables $\vum$ and $\vb$ defined in \eqref{equexplicit} and \eqref{eqvexplicit}.

\begin{remark}
If we rewrite the Hamiltonian \eqref{eq:Hamilt} in the following equivalent form:
\begin{equation}\label{eq:Hpos}
  2\,\H\ =\ \int\left\{\,g\,\eta^2\ -\ g\,d^2\ +\ h\,\vum^2\ +\ h\,\vb^2\ +\ 2\,h\,\vb\,\partial_td\,\right\}\,\ud^2\x,
\end{equation}
one can see that the Hamiltonian density is actually the physical energy density $\mathscr{E}$ if the bottom is static (\ie, if $\partial_td= 0$), but these two quantities are different if the bottom moves. In other words, the Hamiltonian is the energy only if there is no external input of energy into the system. Note also that the Hamiltonian structure of the classical Saint-Venant equations can be recovered substituting $\vb = 0$ into the last Hamiltonian \eqref{eq:Hpos}:
\begin{equation*}
  2\,\H_0\ =\ \int\left\{\,g\,\eta^2\ -\ g\,d^2\ +\ h\,\vum^2\,\right\}\,\ud^2\x,
\end{equation*}
where $\vum = \nab\phim$
\end{remark}


\subsection{Steady solutions}

We consider here the two-dimensional case (\ie, one horizontal dimension) in order to derive a closed form solution for a steady state flow over a general bathymetry. We assume the following upstream conditions at $x\to -\infty$:
\begin{equation*}
  \eta\ \to\ 0, \qquad d\ \to\ d_0, \qquad \um\ \to\ u_0\ \geqslant\ 0.
\end{equation*}
Physically, these conditions mean that far upstream we consider a uniform current over a horizontal bottom. The mass conservation in steady condition yields
\begin{equation*}
  h\,\um\ =\ d_0\, u_0,
\end{equation*}
while the momentum conservation equation becomes
\begin{equation*}
  g\,h\ +\ \half\,\um^2\left[\,1+\,(\partial_xd)^2\,\right]\, =\ g\,d_0\ +\ \half\, u_0^{\,2}.
\end{equation*}

The last two relations yield the following cubic equations for the total water depth (with the dimensionless height $Z=h/d_0>0$ and the Froude number $\Fr=u_0/\sqrt{gd_0} \geqslant 0$)
\begin{equation}\label{eq:steady+}
  G(Z)\ \equiv\ Z^3\ -\ (\,1\, +\, \half\, \Fr^{\,2}\,)\,Z^2\ +\ \half\,\Fr^{\,2}\,[\,1\, +\,(\partial_xd)^2\,]\ =\ 0.
\end{equation}
Note that $G(0) > 0$ for all $\Fr > 0$, $G$ has a maximum at $Z = 0$ and a minimum at $Z = Z_1 = (2 + \Fr^{\,2})/3$. Therefore, \eqref{eq:steady+} has two positive solutions if $G(Z_1) < 0$, one positive solution if $G(Z_1) = 0$, and no positive solutions if $G(Z_1) > 0$. Equation \eqref{eq:steady+} has always a real negative root which  is of no interest for obvious physical reasons.

If  $G(Z_1)<0$, the two positive solutions may be presented as
\begin{equation*}
  Z^+\ =\, \left[\,2\,\sqrt{A/3}\/\cos\!\left(\onethird\arccos\left(-3^{-1/2}BA^{-3/2}\right)-\,\twothirds\pi\right)\right]^{-1},
\end{equation*}
and 
\begin{equation*}
  Z^-\ =\, \left[\,2\,\sqrt{A/3}\/\cos\!\left(\onethird\arccos\left(-3^{-1/2}BA^{-3/2}
\right)\right)\right]^{-1},
\end{equation*}
where
\begin{align*}
  A\ &\equiv\ \frac{1\,+\,2\,\Fr^{-2}}{1\,+\,(\partial_xd)^2}\ \geqslant\ 0, \qquad
  B\ \equiv\ \frac{9\,\Fr^{-2}}{1\,+\,(\partial_xd)^2}\ \geqslant\ 0.
\end{align*}
We note that $Z^- < Z^+$. The root $Z = Z^+$ corresponds to the subcritical regime, while $Z = Z^-$ corresponds to a supercritical regime. For the special case $\Fr = 1$, we have $Z^+ > 1$ and $Z^- < 1$.

If  $G(Z_1) = 0$, for a given Froude number $\Fr$, there is only one absolute value of the slope for which this identity is satisfied, that is
\begin{equation}\label{eq:Froude}
  (\partial_xd)^2\ =\ (\Fr^2-1)^2\,(\Fr^2+8)\,/\,27\,\Fr^2.
\end{equation}
For instance, if $\partial_x d = 0$ then $G(Z_1) = 0$ if and only if $\Fr = 1$.

\begin{remark}
It is straightforward to derive a similar equation for steady solutions to the classical Saint-Venant equations
\begin{equation*}
  Z^3\ -\ (\,1\, +\, \half\, {\Fr}^{\,2}\,)\,Z^2\ +\ \half\,{\Fr}^{\,2}\ =\ 0.
\end{equation*}
The last relation can be also obtained from equation \eqref{eq:steady+} by taking $\partial_x d = 0$. Consequently, we can say that steady solutions to classical Saint-Venant equations do not take into account the bottom slope local variations.
\end{remark}

\begin{figure}
  \centering
  \includegraphics[width=0.58\textwidth]{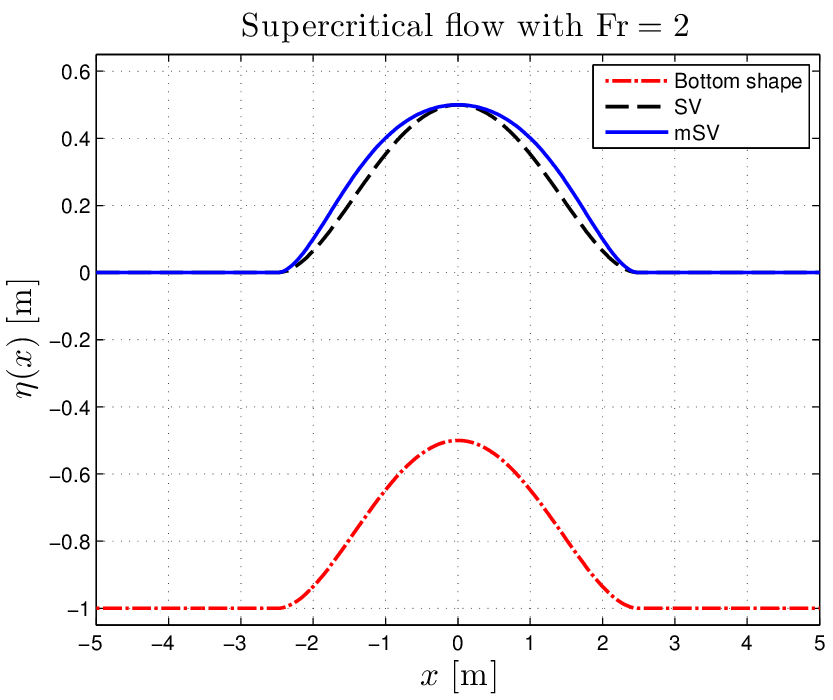}
  \caption{\small\em Supercritical steady state solutions over a bump for the Froude number $\Fr = 2$. Comparison between the classical and modified Saint-Venant equations.}
  \label{fig:steady}
\end{figure}

In order to illustrate the developments made above, we compute a steady flow over a bump. The bottom takes the form
\begin{equation*}
  d(x)\ =\ d_0\ -\ a\,b^{-4}\left(\,x^2\,-\,b^2\,\right)^{\!2} \operatorname{H}(b^2-x^2),
\end{equation*}
where $\operatorname{H}(x)$ is the Heaviside step function \cite{Abramowitz1965}, $a$ and $b$ being the bump amplitude and its half-length, respectively. The values of various parameters are given in Table~\ref{tab:bump}. We consider here for illustrative purposes the supercritical case for the classical and new models. The result are shown on Figure~\ref{fig:steady} where some small differences can be noted with respect to the classical Saint-Venant equations.

\begin{table}
  \centering
  \begin{tabular}{l|c}
  \hline\hline
  \textit{Parameter} & \textit{Value} \\
  \hline\hline
  Gravity acceleration $g$:  & $1\,\mathsf{m\,s^{-2}}$ \\
  Undisturbed water depth $d_0$: & $1\,\mathsf{m}$ \\
  Deformation amplitude $a$: & $0.5\,\mathsf{m}$ \\
  Half-length of the uplift area $b$: & $2.5\,\mathsf{m}$ \\
  Upstream flow speed, $u_0$: & $2.0\,\mathsf{m\,s^{-1}}$ \\
  \hline\hline
  \end{tabular}
  \bigskip
  \caption{\small\em Values of various parameters used for the steady state computation.}
  \label{tab:bump}
\end{table}

\section{Numerical methods}\label{sec:fv}

In this Section we discuss some properties of the system \eqref{eq:mas}, \eqref{eq:qdm} and then, we propose a space discretization procedure based on the finite volume method along with a high-order adaptive time stepping.

\subsection{Hyperbolic structure}

From now on, we consider equations \eqref{eq:mas}, \eqref{eq:qdm} posed in one horizontal space dimension (two-dimensional waves) for simplicity:
\begin{align}
  \partial_t\,h\ +\ \partial_x\,[\,h\,\um\,]\ =&\ 0, \label{eq:mass} \\
  \partial_t\left[\,\um\,-\,\vb\,\partial_x\,d\,\right]\, +\ \partial_x\left[\,g\,\eta\, +\, \half\,\um^2\, +\, \half\, \vb^2\, +\, \vb\,\partial_t\,d\,\right]\, = &\ 0. \label{eq:um}
\end{align}
In order to present the equations in a more suitable conservative form, we will introduce the potential velocity variable $U = \partial_x\phim$. From equation \eqref{eqdvu} it is straightforward to see that $U$ satisfies the relation
\begin{equation*}
  U\ =\ \um\ -\ \vb\,\partial_x\,d,
\end{equation*}
Depth averaged and vertical bottom velocities can be also easily expressed in terms of the potential velocity $U$
\begin{equation*}
  \um\ =\ \frac{U\, -\,(\partial_t\/d)\,(\partial_x\/d)}{1\,+\,(\partial_x\/d)^2}, \qquad
  \vb\ =\ -\frac{\partial_t\/d\,+\,U\,\partial_x\/d}{1\,+\,(\partial_x\/d)^2}.
\end{equation*}
Consequently, using this new variable equations \eqref{eq:mass}, \eqref{eq:um} can be rewritten as a system of conservation laws
\begin{align*}
  \partial_t\,h\ +\ \partial_x\left[\,h\,\frac{U\,-\,(\partial_t\/d)\,(\partial_x\/d)}{1\,+\,(\partial_x\/d)^2}\,\right]\, =&\ 0, \\
  \partial_t\,U\ +\ \partial_x\left[\,g\,(h-d)\, +\, \frac{1}{2}\frac{U^2\,-\,2\,U\,(\partial_t\/d)\,(\partial_x\/d) - (\partial_t\/d)^2}{1\,+\,(\partial_x\/d)^2} \,\right]\, =&\ 0.
\end{align*}
For the sake of simplicity, we rewrite the above system in the following quasilinear vectorial form:
\begin{equation}\label{eq:conslaw}
  \partial_t\,w\ +\ \partial_x\,f(w)\ =\ 0,
\end{equation}
where we introduced the vector of conservative variables $w$ and the advective flux $f(w)$:
\begin{equation*}
w\ = \begin{pmatrix}
        h \\
        U
      \end{pmatrix}, \qquad
  f(w) = \begin{pmatrix}
            \displaystyle{h\frac{U-(\partial_t\/d)(\partial_x\/d)}{1+(\partial_x\/d)^2}} \\
            g(h-d) + \displaystyle{\frac{U^2 - 2U(\partial_x\/d)(\partial_t\/d) - 
            (\partial_t\/d)^2}{2\,[1+(\partial_x\/d)^2]}}
         \end{pmatrix}.
\end{equation*}
The Jacobian matrix of the advective flux $f(w)$ can be easily computed:
\begin{equation*}
  \A(w)\ =\ \pd{\,f(w)}{w}\ =\ \frac{1}{1\/+\/(\partial_x\/d)^2} 
  \begin{bmatrix}
    \displaystyle{U-(\partial_t\/d)\/(\partial_x\/d)} & \displaystyle{h} \\
    g\,(1+(\partial_x\/d)^2) & \displaystyle{U-(\partial_t\/d)\/(\partial_x\/d)}
  \end{bmatrix}\ =\  
  \begin{bmatrix}
    \um & \displaystyle{\frac{h}{1\/+\/(\partial_x\/d)^2}} \\
    g & \um
  \end{bmatrix}.
\end{equation*}
The matrix $\A(w)$ has two distinct eigenvalues:
\begin{equation*}
  \lambda^\pm\ =\ \frac{U\,-\,(\partial_t\/d)\,(\partial_x\/d)}{1\,+\,(\partial_x\/d)^2}\ \pm\ c\ =\ \um\ \pm\ c, \qquad c^2\ \equiv\ \frac{g\,h}{1\,+\,(\partial_x\/d)^2}.
\end{equation*}

\begin{remark}\label{rem:cs}
Physically, the quantity $c$ represents the phase celerity of long gravity waves. In the framework of the Saint-Venant equations, it is well known that $c = \sqrt{gh}$. Both expressions differ by the factor $1/\sqrt{1+(\partial_x\/d)^2}$. In our model, the long waves are slowed down by strong bathymetric variations since fluid particles are constrained to follow the seabed. We note also that a similar factor was previously introduced in \cite{Gobbi1999} to account for steepness in the bathymetry. In our case it appears naturally as a property of the model.
\end{remark}

Right and left eigenvectors coincide with those of the Saint-Venant equations and they are given by the following matrices
\begin{equation*}
R\ =\ \begin{bmatrix}
      -h & h \\
      \sqrt{gh} & \sqrt{gh}
    \end{bmatrix}, \qquad
L\ =\ \frac{1}{2} \begin{bmatrix}
      -h^{-1} & (gh)^{-1/2} \\
      \ h^{-1} & (gh)^{-1/2}
    \end{bmatrix}.
\end{equation*}
Columns of the matrix $R$ constitute eigenvectors corresponding to eigenvalues $\lambda^-$ and $\lambda^+$ respectively. Corresponding left eigenvectors are conventionally written in lines of the matrix $L$.


\subsection{Group velocity}

We would like to compute also the group velocity in the framework of the modified Saint-Venant equations. This quantity is traditionally associated to the wave energy propagation speed \cite{Stoker1958, Dutykh2009b}. Recall, that in the classical linearized shallow water theory, the phase $c$ and group $c_g$ velocities are equal \cite{Stoker1958}:
\begin{equation*}
  c\ =\ \frac{\omega}{k}\ =\ \sqrt{gh}, \qquad 
  c_g\ =\ \od{\,\omega}{k}\ =\ \sqrt{gh},
\end{equation*}
where $\omega = k\sqrt{gh}$ is the dispersion relation for linear long waves, $k$ being the wavenumber and $\omega$ being the angular frequency.

In order to assess the wave energy propagation speed we will consider a quasilinear system of equations composed of mass and energy conservation laws:
\begin{eqnarray*}
  \partial_t\/h\ +\ \partial_x\left[\,h\,\frac{U-(\partial_x\/d)(\partial_t\/d)}{1+(\partial_x\/d)^2}\,\right]\ &=&\ 0, \\
  \partial_t\/E\ +\ \partial_x\left[\,(\/E\/+\/\half\/g\/h^2\/)\frac{U-(\partial_x\/d)(\partial_t\/d)}{1+(\partial_x\/d)^2}\,\right] &=& -\/(g+\gamma)\,h\,\partial_t\/d,
\end{eqnarray*}
where $\gamma$ is defined in \eqref{eq:acc} and $E$ is the total energy considered already above \eqref{eq:energy}:
\begin{equation*}
  E\ =\ h\,\frac{\um^2 + \vb^2}{2}\ +\ \frac{g\,(\eta^2 - d^2)}{2}\ =\ \frac{h}{2}\,\frac{U^2 + (\partial_t\/d)^2}{1\,+\,(\partial_x\/d)^2}\ +\ \frac{g\,(h^2 - 2hd)}{2}.
\end{equation*}
The last formula can be inverted to express the potential velocity in terms of the wave energy:
\begin{equation*}
  U^2\ =\ [\,1\,+\,(\partial_x\/d)^2\,]\left(\,\frac{2\/E}{h}\, -\, g\/h\, +\, 2\/g\/d\,\right)\ -\ (\partial_t\/d)^2.
\end{equation*}
In the spirit of computations performed in the previous section, we compute the Jacobian matrix $\J$ of the mass-energy advection operator:
\begin{equation*}
  \J\ =\ \frac{1}{1\,+\,(\partial_x\/d)^2}
\begin{bmatrix}
    U\, -\, (\partial_x\/d)(\partial_t\/d)\, +\, h\/\displaystyle{\pd{\,U}{h}} & h\displaystyle{\pd{\,U}{E}} \\
    g\/h\/[U\/ -\/ (\partial_x\/d)(\partial_t\/d)]\, +\, (E\/+\/\half\/g\/h^2)
    \displaystyle{\pd{\,U}{h}} & U\,-\,(\partial_x\/d)(\partial_t\/d)\, +\, (E+\half\/g\/h^2)
    \displaystyle{\pd{\,U}{E}}
\end{bmatrix},
\end{equation*}
where partial derivatives are given here:
\begin{equation*}
  \pd{\,U}{h}\ =\ -\,[\,1\,+\,(\partial_x\/d)^2\,]\,\frac{g\/h^2\,+\,2\/E}{2\,h^2\,U}, \qquad
  \pd{\,U}{E}\ =\ \frac{1\,+\,(\partial_x\/d)^2}{h\,U}.
\end{equation*}
Computation of the Jacobian $\J$ eigenvalues leads the following expression for the \emph{group velocity} of the modified Saint-Venant equations:
\begin{equation*}
  c_g^{\,2}\ =\ \frac{g\,h}{1\,+\,(\partial_x\/d)^2}\frac{U\,-\,(\partial_x\/d)(\partial_t\/d)}{U}.
\end{equation*}
The last formula is very interesting. It means that in the moving bottom case, the group velocity $c_g$ is modified and does not coincide anymore with the phase velocity $c^2 = gh[1+(\partial_x\/d)^2]^{-1}$. This fact represents another new and non-classical feature of the modified Saint-Venant equations. The relative difference between phase and group velocities squared is
\begin{align*}
  \frac{c^2\,-\,c_g^{\,2}}{c^2}\ =\ \frac{(\partial_x\/d)\,(\partial_t\/d)}{U},
\end{align*}
which is not necessarily always positive. When it is negative, the energy is injected into the system at a higher rate than can be spread, thus leading to energy accumulation and possibly favoring the breaking events.


\subsection{Finite volume scheme}

Let us fix a partition of $\R$ into cells (or finite volumes) $\C_i = [x_{i-\frac12}, x_{i+\frac12}]$ with cell center $x_i = \half(x_{i-\frac12} + x_{i+\frac12})$, $i\in\Z$. Let $\Delta x_i$ denotes the length of the cell $\C_i$. Without any loss of generality we assume the partition to be uniform, \ie, $\Delta x_i = \Delta x$, $\forall i\in\Z$. The solution $w(x,t)$ is  approximated by discrete values and, in order to do so, we introduce the cell average of $w$ over the cell $\C_i$, \ie,
\begin{equation*}
  \wb_i(t)\ \equiv\ \frac{1}{\Delta h} \int_{\C_i} w(x,t)\,\ud\/x.
\end{equation*}
A simple integration of \eqref{eq:conslaw} over the cell $\C_i$ leads the following exact relation
\begin{equation*}
  \od{\,\wb_i}{t}\ +\ \frac{f\left(w(x_{i+\frac12},t)\right)\, -\ f\left(w(x_{i-\frac12},t)\right)}{\Delta x}\ =\ 0.
\end{equation*}
Since the discrete solution is discontinuous at the cell interface $x_{i+\frac12}$, $i\in\Z$, the heart of the matter in the finite volume method is to replace the flux through cell faces by the so-called numerical flux function
\begin{equation*}
  f\left(w(x_{i\pm\frac12},t)\right)\, \approx\ \F_{i\pm\frac12}\left(\wb_{i\pm\frac12}^{L}, \wb_{i\pm\frac12}^{R}\right),
\end{equation*}
where $\wb_{i\pm\frac12}^{L,R}$ are reconstructions of conservative variables $\wb$ from left and right sides of each cell interface \cite{Barth2004, Leer2006}. The reconstruction procedure employed in the present study is described below.

In order to discretise the advective flux $f(w)$, we use the so-called FVCF scheme \cite{Ghidaglia2001}
\begin{equation*}
  \F(v,w)\ =\ \frac{f(v)\,+\,f(w)}{2}\ -\ \S(v,w)\,\frac{f(w)\,-\,f(v)}{2}.
\end{equation*}
The first part of the numerical flux $\F(v,w)$ is centered, while the second part is the upwinding introduced according to local waves propagation directions
\begin{equation*}
  \S(v,w)\, =\, \sign\left(\A\left(\frac{v+w}{2}\right)\right), \quad \sign(\A)\, =\, R\scal\diag(s^-, s^+)\scal L, \quad s^\pm\, \equiv\, \sign(\lambda^\pm).
\end{equation*}
After some simple algebraic computations one can find expressions for sign matrix $\S$ coefficients
\begin{equation*}
  \S\ =\ \frac{1}{2}\begin{bmatrix}
   		\,(s^++s^-)\, & \,(s^+-s^-)\sqrt{h/g}\, \\
   		\,(s^+-s^-)\sqrt{g/h}\, & \,(s^++s^-)\,
      \end{bmatrix},
\end{equation*}
all coefficients being evaluated at the average state of left and right face values.

Taking into account the developments presented above, the semi-discrete scheme takes the form
\begin{equation}\label{eq:si1}
  \od{\,\wb_i}{\/t}\ +\ \frac{\F_{i+\frac12}\left(\wb_{i+\frac12}^{L},\wb_{i+\frac12}^{R}\right) - \F_{i-\frac12}\left(\wb_{i-\frac12}^{L},\wb_{i-\frac12}^{R}\right)}{\Delta x}\ =\ 0.
\end{equation}
The discretization in time of the last system of ODEs is discussed in Section~\ref{sec:timestep}. Meanwhile, we present the employed reconstruction procedure. The method presented below has already been successfully applied to dispersive problems \cite{Dutykh2010e, Dutykh2011a}.


\subsection{High-order reconstruction}

In order to obtain a higher-order scheme in space, we need to replace the piecewise constant data by a piecewise polynomial representation. This goal is achieved by various so-called reconstruction procedures, such as MUSCL TVD \cite{Kolgan1975, Leer1979}, UNO \cite{HaOs}, ENO \cite{Harten1989}, WENO \cite{Xing2005} and many others. In our previous study on Boussinesq-type equations \cite{Dutykh2011e}, the UNO2 scheme showed a good performance with low dissipation in realistic propagation and runup simulations. Consequently, we retain this scheme for the discretization of the modified Saint-Venant equations.

\begin{remark}
In TVD schemes, the numerical operator is required (by definition) not to increase the total variation of the numerical solution at each time step. It follows that the value of an isolated maximum may only decrease in time which is not a good property for the simulation of coherent structures such as solitary waves. The non-oscillatory UNO2 scheme, employed in our study, is only required to diminish the \emph{number} of local extrema in the numerical solution. Unlike TVD schemes, UNO schemes are not constrained to damp the values of each local extremum at every time step.
\end{remark}

The main idea of the UNO2 scheme is to construct a non-oscillatory piecewise-parabolic interpolant $Q(x)$ to a piecewise smooth function $w(x)$ (see \cite{HaOs} for more details). On each segment containing the face $x_{i+\frac12} \in [x_i, x_{i+1}]$, the function $Q(x) = q_{i+\frac12}(x)$ is locally a quadratic polynomial $q_{i+\frac12}(x)$ and wherever $w(x)$ is smooth we have
\begin{equation*}
  Q(x)\ -\ w(x)\ =\ \O(\Delta x^3), \qquad
  \od{\,Q}{\/x}(x\pm 0)\ -\ \od{\,w}{\/x}\ =\ \O(\Delta x^2).
\end{equation*}
Also $Q(x)$ should be non-oscillatory in the sense that the number of its local extrema does not exceed that of $w(x)$. Since $q_{i+\frac12}(x_i)=\wb_i$ and $q_{i+\frac12}(x_{i+1}) = \wb_{i+1}$, it can be written in the form
\begin{equation*}
  q_{i+\frac12}(x)\ =\ \wb_i\ +\ d_{i+\frac12}(w)\,\frac{x - x_i}{\Delta x}\ +\ \half D_{i+\frac12}w\cdot\frac{(x-x_i)(x-x_{i+1})}{\Delta x^2},
\end{equation*}
where $d_{i+\frac12}(w)\equiv\wb_{i+1}-\wb_{i}$ and $D_{i+\frac12}v$ is closely related to the second derivative of the interpolant since $D_{i+\frac12}v = \Delta x^2 q''_{i+\frac12}(x)$. The polynomial $q_{i+\frac12}(x)$ is chosen to be one the least oscillatory between two candidates interpolating $w(x)$ at $(x_{i-1}, x_i, x_{i+1})$ and $(x_i, x_{i+1}, x_{i+2})$. This requirement leads to the following choice of $D_{i+\frac12}v$
\begin{equation*}
  D_{i+\frac12}w := \minmod\bigl(D_iw, D_{i+1}w\bigr), \quad
  D_iw = \wb_{i+1} - 2\wb_i + \wb_{i-1}, \quad
  D_{i+1}w = \wb_{i+2} - 2\wb_{i+1} + \wb_i,
\end{equation*}
and $\minmod(x,y)$ is the usual min mod function defined as:
\begin{equation*}
  \minmod(x,y)\ =\ \half\,[\,\sign(x)\,+\,\sign(y)\,]\times\min(|x|,|y|).
\end{equation*}

To achieve the second order $\O(\Delta x^2)$ accuracy, it is sufficient to consider piecewise linear reconstructions in each cell. Let $L(x)$ denote this approximately reconstructed function which can be written in the form
\begin{equation*}
  L(x)\ =\ \wb_i\ +\ s_i\,(x-x_i)\,/\,\Delta x, \qquad x_{i-\frac12}\,\leqslant\,x\, \leqslant x_{i+\frac12}.
\end{equation*}
To make $L(x)$ a non-oscillatory approximation, we use the parabolic interpolation $Q(x)$ constructed below to estimate the slopes $s_i$ within each cell
\begin{equation*}
  s_i\ =\ \Delta x\times\minmod\!\left(\left.\od{\,Q}{\/x}\right|_{x=x_i^-} , \left.\od{\,Q}{\/x}\right|_{x=x_i^+}\right).
\end{equation*}
In other words, the solution is reconstructed on the cells while the solution gradient is estimated on the dual mesh as it is often performed in more modern schemes \cite{Barth2004}. A brief summary of the UNO2 reconstruction can be also found in \cite{Dutykh2011e}.


\subsection{Time stepping}\label{sec:timestep}

We rewrite the semi-discrete scheme \eqref{eq:si1} as a system of ODEs:
\begin{equation*}
  \partial_t\,\wb\ =\ \L (\wb, t), \qquad \wb(0)\ =\ \wb_0.
\end{equation*}
In order to solve numerically the last system of equations, we apply the Bogacki--Shampine method \cite{Bogacki1989}. It is a third-order Runge--Kutta scheme with four stages. It has an embedded second-order method which is used to estimate the local error and, thus, to adapt the time step size. Moreover, the Bogacki--Shampine method enjoys the First Same As Last (FSAL) property so that it needs three function evaluations per step. This method is also implemented in the \texttt{ode23} function in {\sc Matlab} \cite{Shampine1997}. A step of the Bogacki--Shampine method is given by
\begin{align*}
  k_1\ =&\ \L(\wb^{(n)},t_n), \\
  k_2\ =&\ \L(\wb^{(n)}+\half \Delta t_n k_1, t_n + \half\Delta t), \\
  k_3\ =&\ \L(\wb^{(n)})+\frth \Delta t_n k_2, t_n + \frth\Delta t), \\
  \wb^{(n+1)}\ =&\ \wb^{(n)}\ +\ \Delta t_n\left(\textstyle{2\over9}k_1 + \textstyle{1\over3}k_2 + \textstyle{4\over9}k_3\right), \\
  k_4\ =&\ \L(\wb^{(n+1)}, t_n + \Delta t_n), \\
  \wb_2^{(n+1)}\ =&\ \wb^{(n)}\ +\ \Delta t_n\left(\textstyle{4\over{24}}k_1 + \textstyle{1\over4}k_2 + \textstyle{1\over3} k_3 + \textstyle{1\over8}k_4\right).
\end{align*}
Here $\wb^{(n)}\approx\wb(t_n)$, $\Delta t$ is the time step and $\wb_2^{(n+1)}$ is a second-order approximation to the solution $\wb(t_{n+1})$, so the difference between $\wb^{(n+1)}$ and $\wb_2^{(n+1)}$ gives an estimation of the local error. The FSAL property consists in the fact that $k_4$ is equal to $k_1$ in the next time step, thus saving one function evaluation.

If the new time step $\Delta t_{n+1}$ is given by $\Delta t_{n+1} = \rho_n\Delta t_n$, then according to H211b digital filter approach \cite{Soderlind2003, Soderlind2006}, the proportionality factor $\rho_n$ is given by:
\begin{equation}\label{eq:tadapt}
  \rho_n\ =\,\left(\frac{\delta}{\epsilon_n}\right)^{\!\beta_1}\left(\frac{\delta}{\epsilon_{n-1}}\right)^{\!\beta_2}\,\rho_{n-1}^{\,-\alpha},
\end{equation}
where $\epsilon_n$ is a local error estimation at time step $t_n$ and constants $\beta_1$, $\beta_2$ and $\alpha$ are defined as
\begin{equation*}
  \alpha\ =\ 1\,/\,4, \qquad \beta_1\ =\ \beta_2\ =\ 1\,/\,4\,p.
\end{equation*}
The parameter $p$ is the order of the scheme ($p=3$ in our case).

\begin{remark}
The adaptive strategy \eqref{eq:tadapt} can be further improved if we smooth the factor $\rho_n$ before computing the next time step $\Delta t_{n+1}$
\begin{equation*}
  \Delta t_{n+1}\ =\ \hat{\rho}_n\,\Delta t_n, \qquad \hat{\rho}_n\ =\ \omega(\rho_n).
\end{equation*}
The function $\omega(\rho)$ is called \emph{the time step limiter} and should be smooth, monotonically increasing and should satisfy the following conditions
\begin{equation*}
  \omega(0)\ <\ 1, \qquad \omega(+\infty)\ >\ 1, \qquad \omega(1)\ =\ 1, \omega'(1)\ =\ 1.
\end{equation*}
One possible choice is suggested in \cite{Soderlind2006}:
\begin{equation*}
  \omega(\rho)\ =\ 1\ +\ \kappa\/\arctan\!\left(\frac{\rho-1}{\kappa}\right).
\end{equation*}
In our computations the parameter $\kappa$ is set to 1.
\end{remark}


\section{Numerical results}\label{sec:numres}

The numerical scheme presented above has already been validated in several studies even in the case of dispersive waves \cite{Dutykh2011e, Dutykh2010e}. Consequently, we do not present here the standard convergence tests which can be found in references cited above. In the present Section we show numerical results which illustrate some properties of modified Saint-Venant equations with respect to their classical counterpart. In the sequel we consider only one-dimensional case for simplicity. The physical domain will be also limited by wall-boundary conditions. Other types of boundary conditions obviously could also be considered.


\subsection{Wave propagation over oscillatory bottom}

We begin the exposition of numerical results by presenting a simple test-case of a wave propagating over a static but highly oscillatory bottom. Let us consider a one-dimensional physical domain $[-10, 10]$ which is discretized into $N = 350$ equal control volumes. The tolerance parameter $\delta$ in the time stepping algorithm is chosen to be $10^{-4}$. The initial condition will be simply a bump localized near the center $x=0$ and posed on the free surface with initial zero velocity field
\begin{equation*}
  \eta_0(x)\ =\ b \sech^2(\kappa x), \qquad u_0(x)\ =\ 0.
\end{equation*}
The bottom is given analytically by the function
\begin{equation*}
  d(x)\ =\ d_0\ +\ a\/\sin(kx).
\end{equation*}
In other words, the bathymetry function $d(x)$ consists of uniform level $d_0$ which is perturbed by uniform oscillations of amplitude $a$. Since the bathymetry is static, the governing equations \eqref{eq:mass} and \eqref{eq:um} are simplified at some point.

Hereafter, we fix two wavenumbers $k_1$ and $k_2$ ($k_1<k_2$) and perform a comparison between numerical solutions to the classical and modified Saint-Venant equations. The main idea behind this comparison is to show the similarity between two solutions for gentle bottoms and, correspondingly, to highlight the differences for stronger gradients. The values of various physical parameters used in numerical simulations are given in Table~\ref{tab:oscil}.

\begin{table}
  \centering
  \begin{tabular}{l|c}
  \hline\hline
  \textit{Parameter} & \textit{Value} \\
  \hline\hline
  Initial wavenumber $\kappa$:   & $1\,\mathsf{m^{-1}}$ \\  
  Gravity acceleration $g$:      & $1\,\mathsf{m\, s^{-2}}$  \\
  Final simulation time $T$:     & $24\,\mathsf{s}$ \\
  Initial wave amplitude $b$:    & $0.2\,\mathsf{m}$ \\
  Undisturbed water depth $d_0$: & $1\,\mathsf{m}$ \\
  Bathymetry oscillation amplitude $a$: & $0.1\,\mathsf{m}$ \\
  Low bathymetry oscillation wavelength $k_1$:  & $2\,\mathsf{m^{-1}}$ \\
  High bathymetry oscillation wavelength $k_2$: & $6\,\mathsf{m^{-1}}$ \\
  \hline\hline
  \end{tabular}
  \bigskip
  \caption{\small\em Values of various parameters used for the wave propagation over an oscillatory bottom test-case.}
  \label{tab:oscil}
\end{table}

\begin{figure}
  \centering
  \subfigure[\it Low oscillations, $k_1$]{\includegraphics[width=0.49\textwidth]%
  {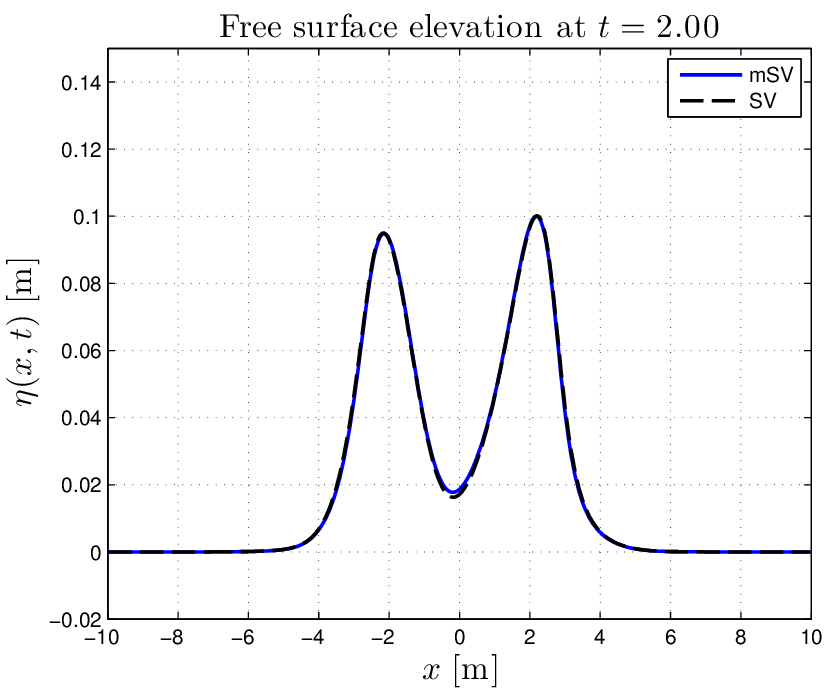}}
  \subfigure[\it High oscillations, $k_2$]{\includegraphics[width=0.49\textwidth]%
  {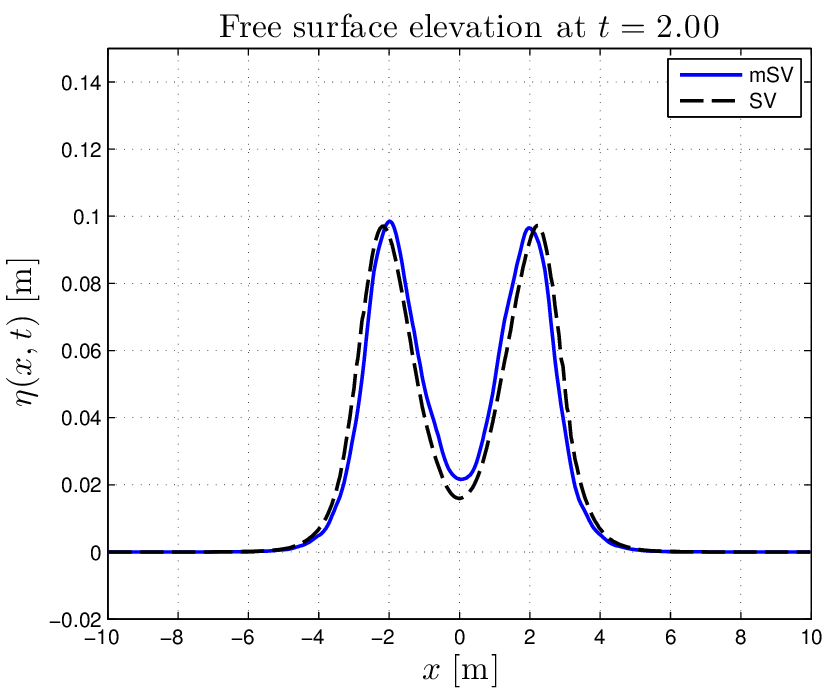}}
  \caption{\small\em Wave propagation over an oscillatory bottom, $t = 2\,\mathsf{s}$.}
  \label{fig:oscilt2}
\end{figure}

\begin{figure}
  \centering
  \subfigure[\it Low oscillations, $k_1$]{\includegraphics[width=0.49\textwidth]%
  {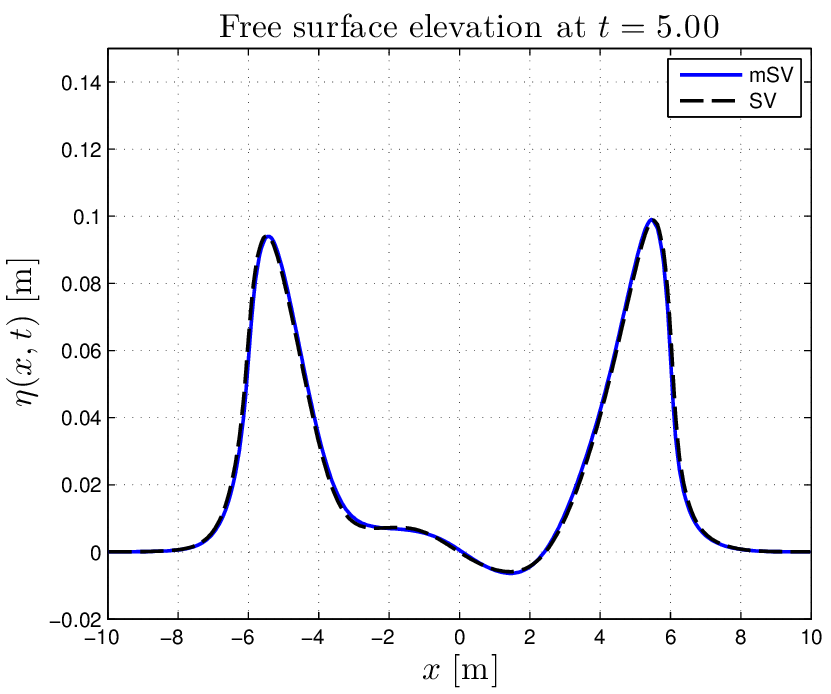}}
  \subfigure[\it High oscillations, $k_2$]{\includegraphics[width=0.49\textwidth]%
  {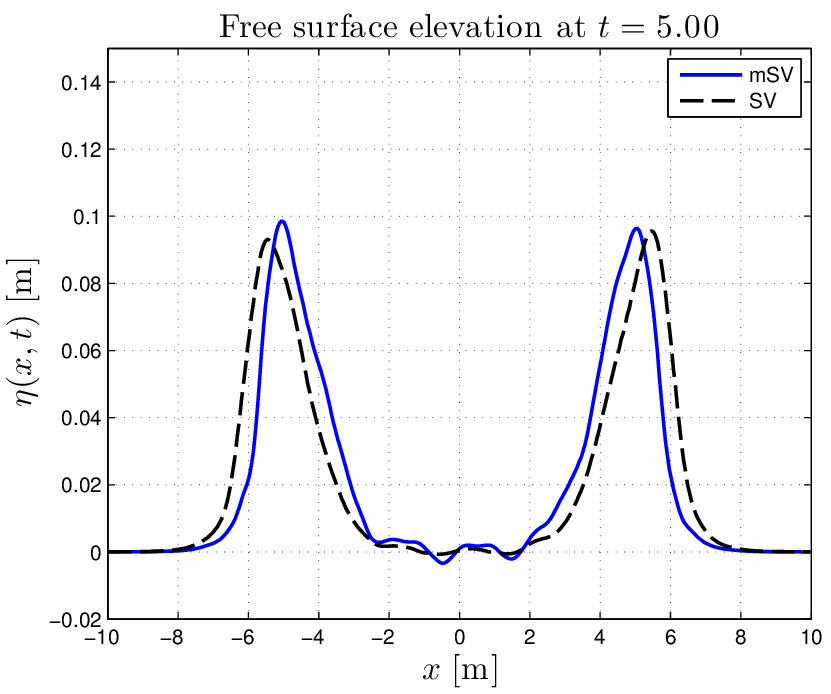}}
  \caption{\small\em Wave propagation over an oscillatory bottom, $t = 5\,\mathsf{s}$.}
  \label{fig:oscilt5}
\end{figure}

\begin{figure}
  \centering
  \subfigure[\it Low oscillations, $k_1$]{\includegraphics[width=0.49\textwidth]%
  {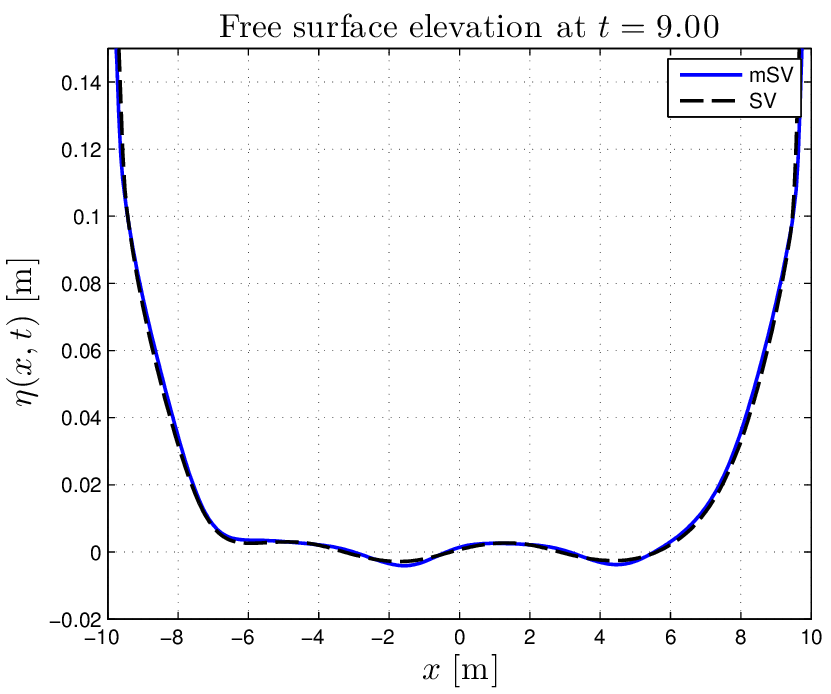}}
  \subfigure[\it High oscillations, $k_2$]{\includegraphics[width=0.49\textwidth]%
  {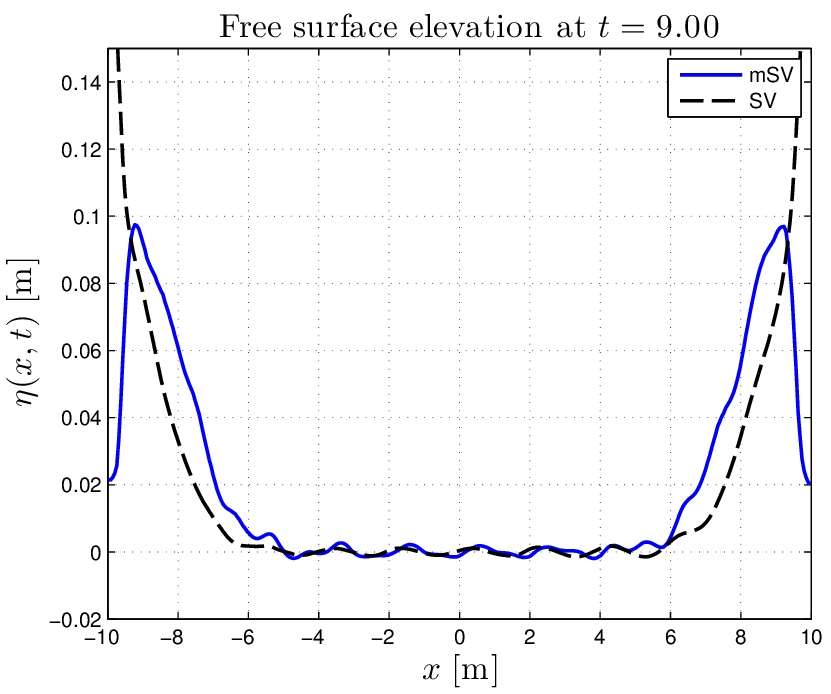}}
  \caption{\small\em Wave propagation over an oscillatory bottom, $t = 9\,\mathsf{s}$.}
  \label{fig:oscilt9}
\end{figure}

\begin{figure}
  \centering
  \subfigure[\it Low oscillations, $k_1$]{\includegraphics[width=0.49\textwidth]%
  {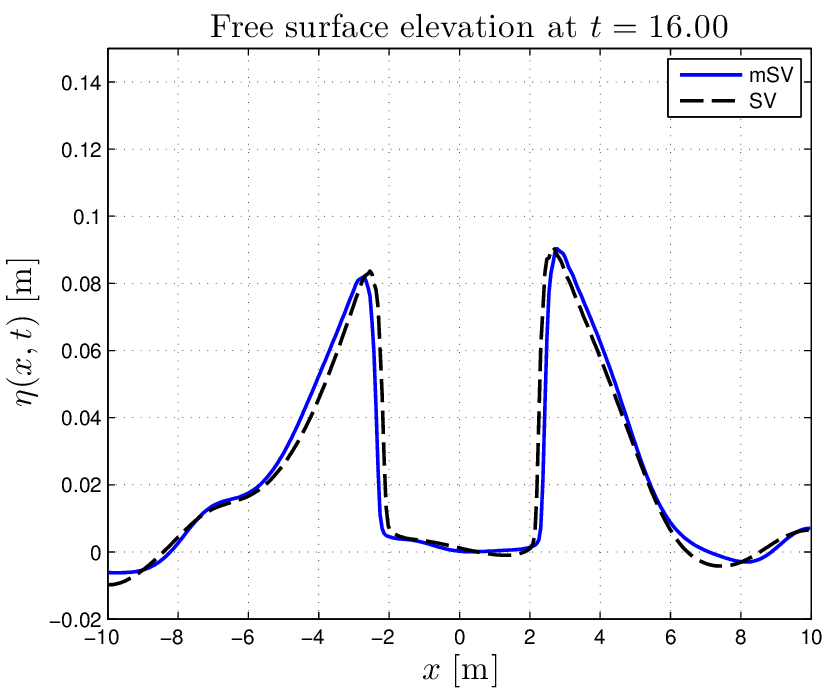}}
  \subfigure[\it High oscillations, $k_2$]{\includegraphics[width=0.49\textwidth]%
  {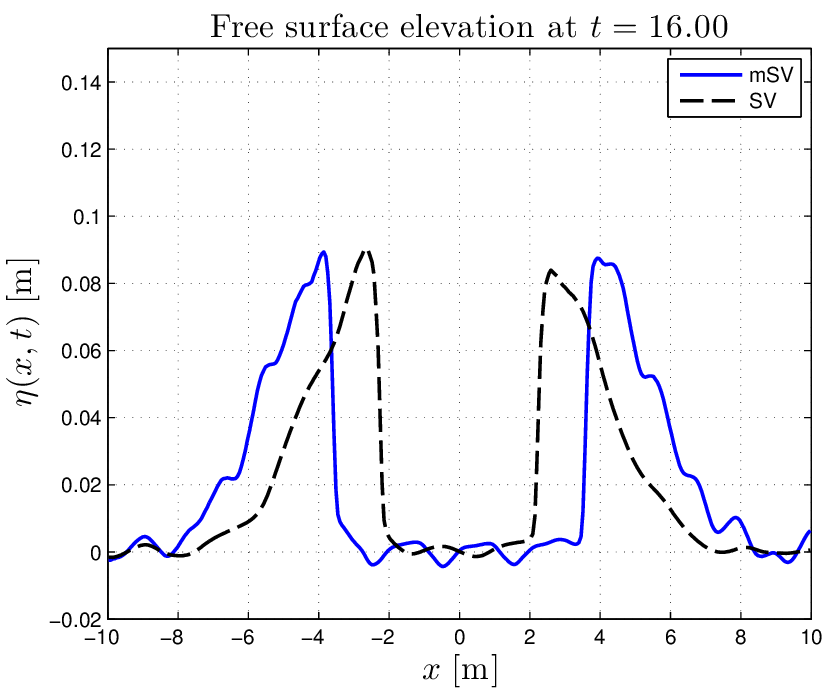}}
  \caption{\small\em Wave propagation over an oscillatory bottom, $t = 16\,\mathsf{s}$.}
  \label{fig:oscilt16}
\end{figure}

\begin{figure}
  \centering
  \subfigure[\it Low oscillations, $k_1$]{\includegraphics[width=0.49\textwidth]%
  {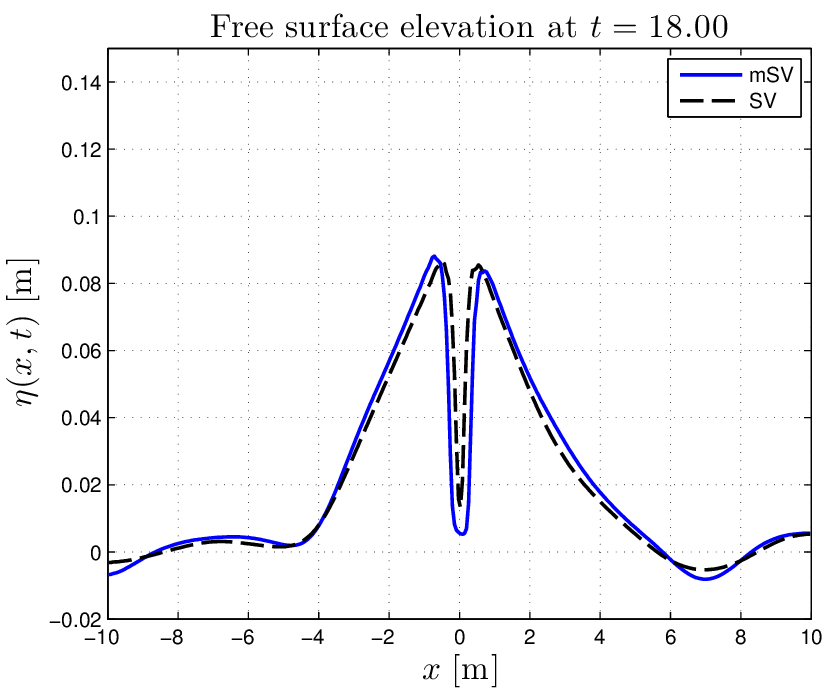}}
  \subfigure[\it High oscillations, $k_2$]{\includegraphics[width=0.49\textwidth]%
  {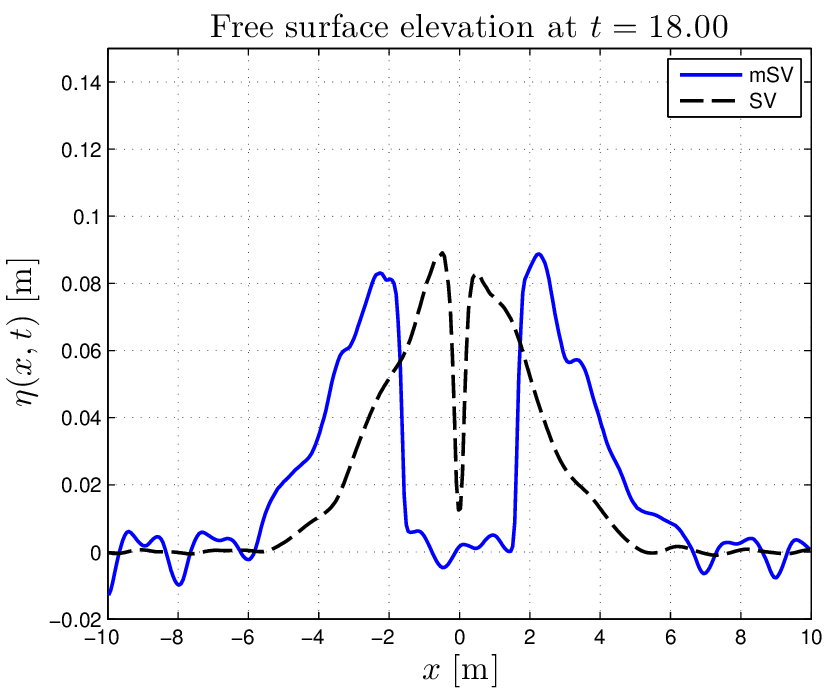}}
  \caption{\small\em Wave propagation over an oscillatory bottom, $t = 18\,\mathsf{s}$.}
  \label{fig:oscilt18}
\end{figure}

\begin{figure}
  \centering
  \subfigure[\it Low oscillations, $k_1$]{\includegraphics[width=0.49\textwidth]%
  {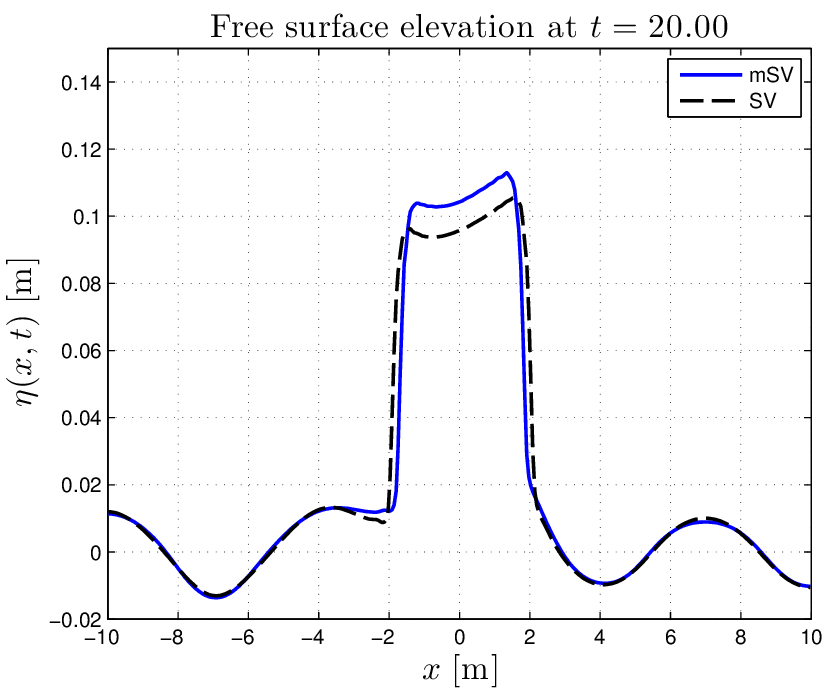}}
  \subfigure[\it High oscillations, $k_2$]{\includegraphics[width=0.49\textwidth]%
  {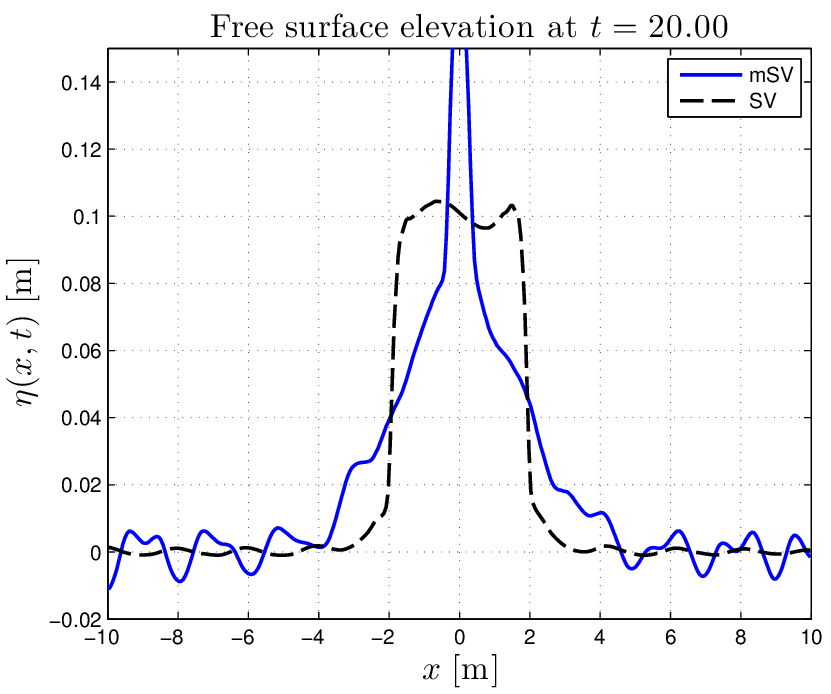}}
  \caption{\small\em Wave propagation over an oscillatory bottom, $t = 20\,\mathsf{s}$.}
  \label{fig:oscilt20}
\end{figure}

\begin{figure}
  \centering
  \subfigure[\it Low oscillations ($k_1$)]{\includegraphics[width=0.49\textwidth]%
  {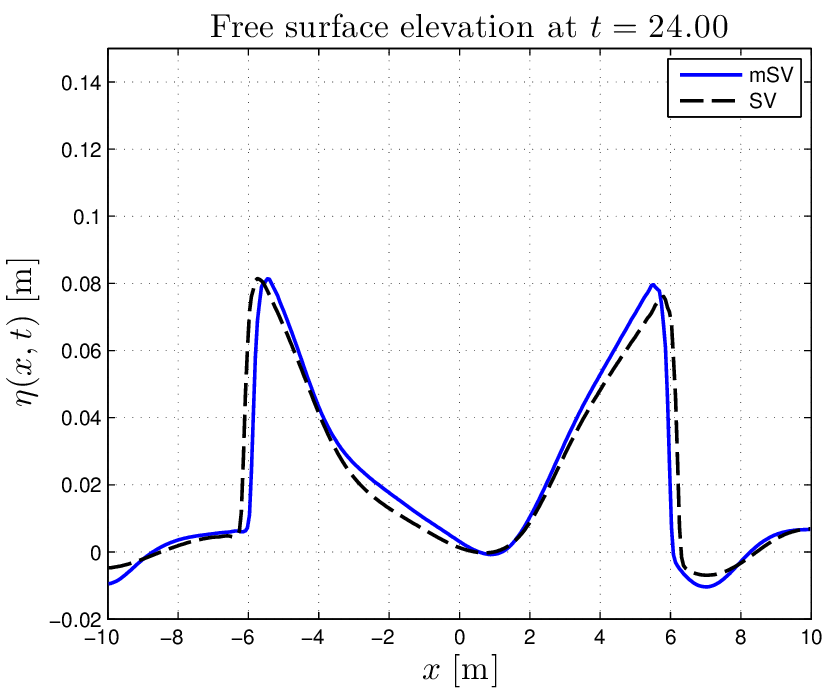}}
  \subfigure[\it High oscillations ($k_2$)]{\includegraphics[width=0.49\textwidth]%
  {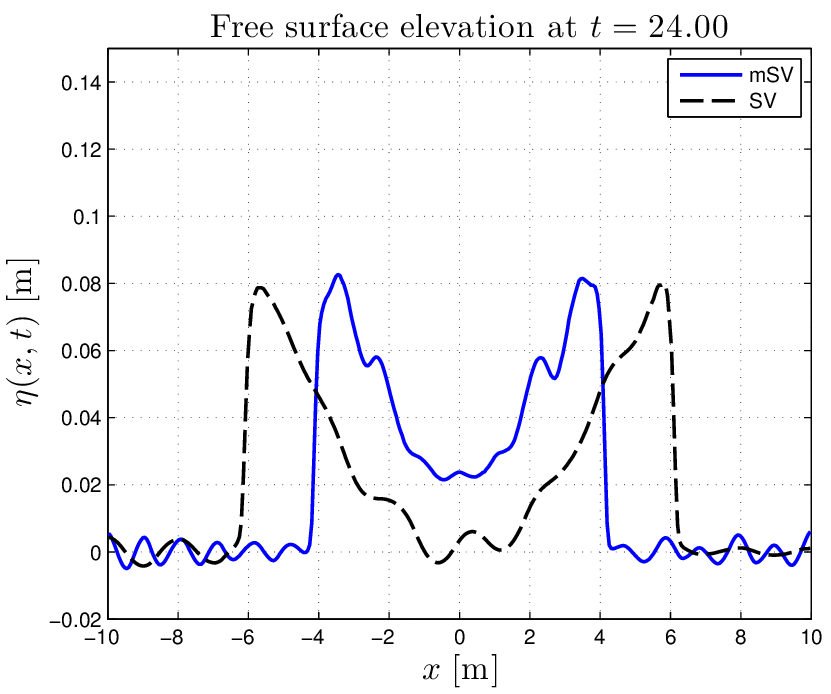}}
  \caption{\small\em Wave propagation over an oscillatory bottom, $t = 24\,\,\mathsf{s}$.}
  \label{fig:oscilt24}
\end{figure}

\begin{figure}
  \centering
  \subfigure[\it Low oscillations ($k_1$)]{\includegraphics[width=0.49\textwidth]%
  {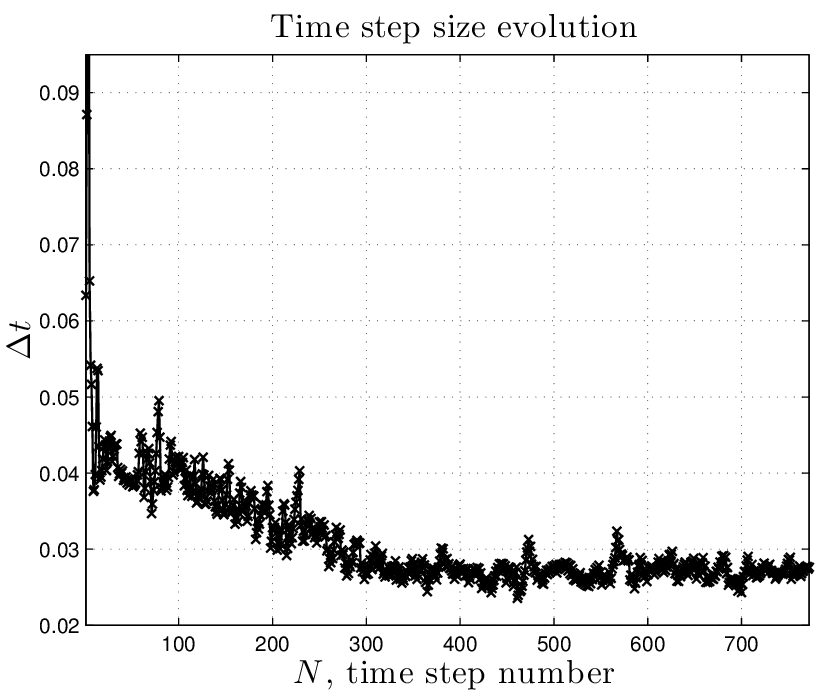}}
  \subfigure[\it High oscillations ($k_2$)]{\includegraphics[width=0.49\textwidth]%
  {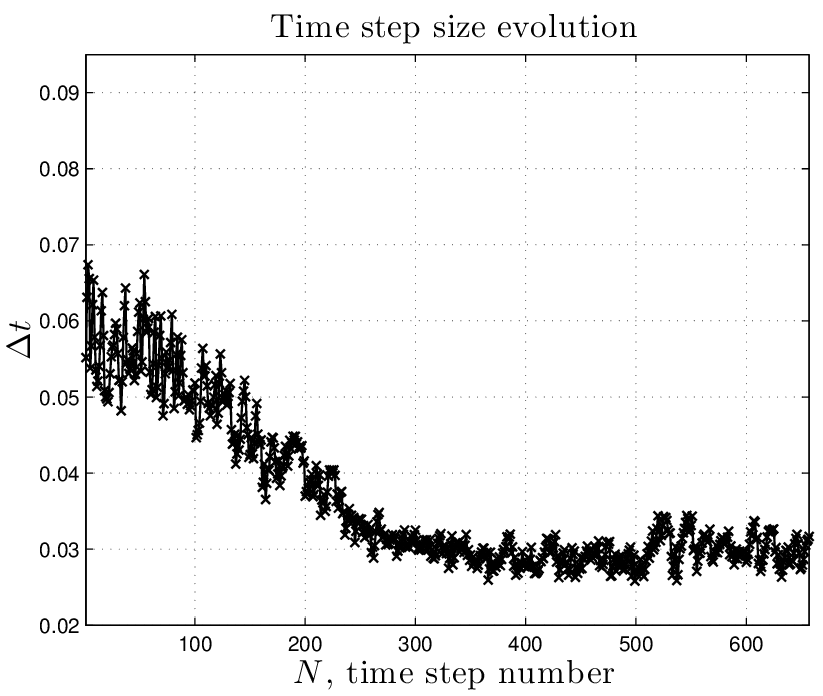}}
  \caption{\small\em Local time step size during the simulation of a wave propagating over an oscillatory bottom test case.}
  \label{fig:timestep}
\end{figure}

Several snapshots of the free surface elevation during the wave propagation test-case are presented on Figures \ref{fig:oscilt2} -- \ref{fig:oscilt24}. The left image refers to the gentle bottom gradient case ($k_1 = 2$) while the right image corresponds to the oscillating bottom ($k_2 = 6$). Everywhere, the solid blue line represents a solution to the \acl{msv} equations, while the dotted black line refers to the classical solution. Numerical results on left images indicate that both models give very similar results when bathymetry gradients are gentle. Two solutions are almost indistinguishable to grahic resolutions, especially at the beginning. However, some divergences are accumulated with time. At the end of the simulation some differences become to be visible to the graphic resolution. On the other hand, numerical solutions on right images are substantially different from first instants of the wave propagation. In accordance with theoretical predictions (see Remark~\ref{rem:cs}), the wave in \acs{msv} equations propagates with speed effectively reduced by bottom oscillations. This fact explains a certain lag between two numerical solutions in the highly oscillating case. We note that the wave shape is also different in classical and improved equations. Finally, on Figure \ref{fig:timestep} we show the evolution of the local time step during the simulation. It can be easily seen that the time adaptation algorithm very quickly finds the optimal value of the time step which is then maintained during the whole simulation. This observation is even more flagrant on the right image corresponding to the highly oscillating case.

\subsection{Wave generation by sudden bottom uplift}

We continue to investigate various properties of the modified Saint-Venant equations. In this section, we present a simple test-case which involves the bottom motion. More precisely, we will investigate two cases of slow and fast uplifts of a portion of bottom. This simple situation has some important implications to tsunami genesis problems \cite{Hammack, Todo, ddk}.

The physical domain and discretization parameters are inherited from the last section. The bottom is given by the following function:
\begin{equation*}
  d(x,t)\ =\ d_0\ -\ a\,T(t)\operatorname{H}(b^2-x^2)\left[\left(\frac{x}{b}\right)^2 - 1\right]^2, \qquad T(t)\ =\ 1\ -\ \mathrm{e}^{-\alpha t},
\end{equation*}
where $\operatorname{H}(x)$ is the Heaviside step function \cite{Abramowitz1965}, $a$ is the deformation amplitude and $b$ is the half-length of the uplifting sea floor area. The function $T(t)$ provides us a complete information on the dynamics of the bottom motion. In tsunami wave literature, it is called a {\em dynamic scenario} \cite{Hammack, Dutykh2006, Kervella2007}. Obviously, other choices of the time dependence are possible. Initially the free surface is undisturbed and the velocity field is taken to be identically zero. The values of various parameter are given in Table~\ref{tab:uplift}.

\begin{table}
  \centering
  \begin{tabular}{l|c}
  \hline\hline
  \textit{Parameter} & \textit{Value} \\
  \hline\hline
  Slow uplift parameter $\alpha_1$:    & $2.0\,\mathsf{s^{-1}}$ \\
  Fast uplift parameter $\alpha_2$:    & $12.0\,\mathsf{s^{-1}}$ \\
  Gravity acceleration $g$:            & $1\,\mathsf{m\,s^{-2}}$ \\
  Final simulation time $T$:           & $5\,\mathsf{s}$ \\
  Undisturbed water depth $d_0$:       & $1\,\mathsf{m}$ \\
  Deformation amplitude $a$:           & $0.25\,\mathsf{m}$ \\
  Half-length of the uplift area $b$:  & $2.5\,\mathsf{m}$ \\
  \hline\hline
  \end{tabular}
  \bigskip
  \caption{\small\em Values of various parameters used for the wave generation by a moving bottom.}
  \label{tab:uplift}
\end{table}

\begin{figure}
  \centering
  \subfigure[$t = 0.5\,\mathsf{s}$]{\includegraphics[width=0.49\textwidth]%
  {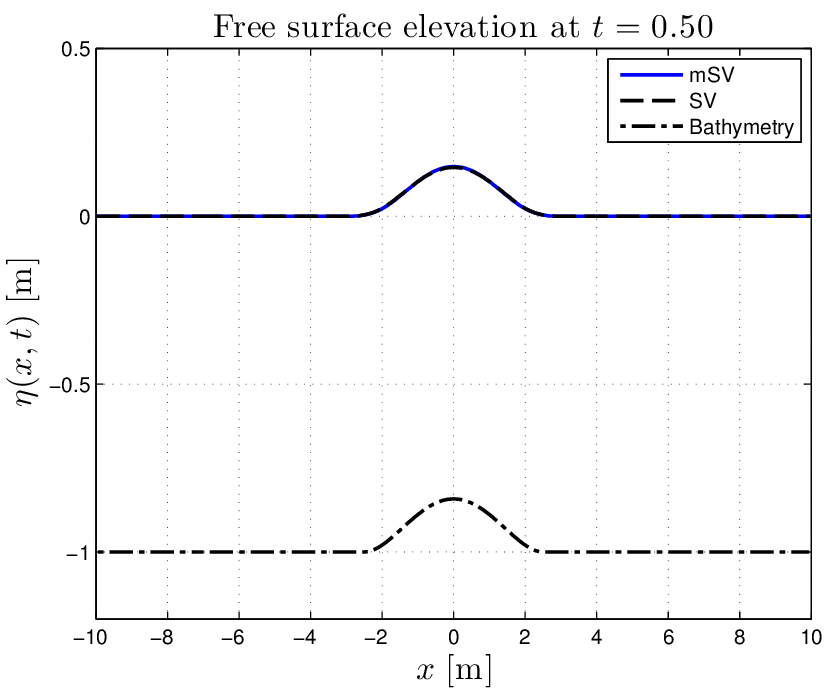}}
  \subfigure[$t = 1.0\,\mathsf{s}$]{\includegraphics[width=0.49\textwidth]%
  {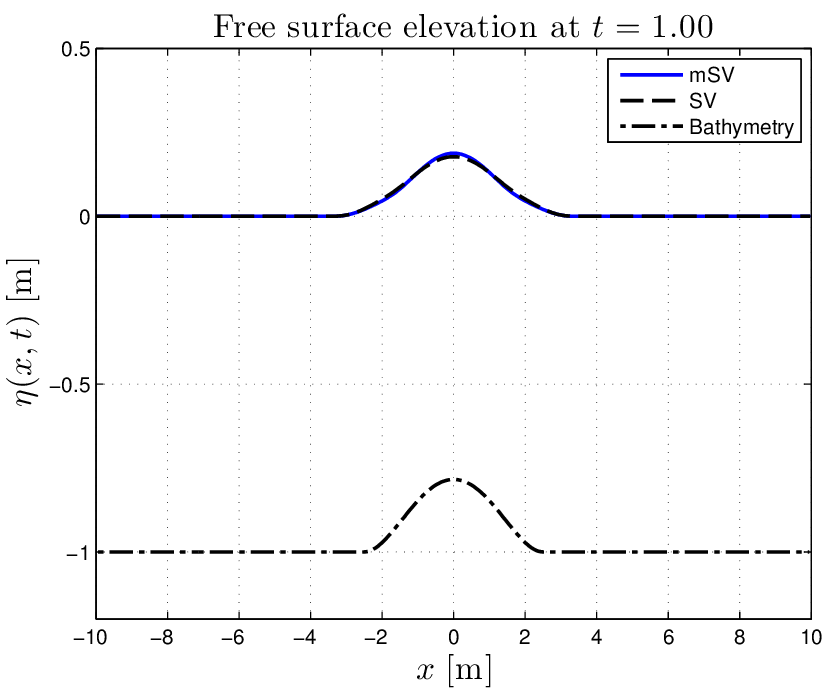}}
  \caption{\small\em Slow bottom uplift test-case ($\alpha_1 = 2$).}
  \label{fig:a2t50-100}
\end{figure}

\begin{figure}
  \centering
  \subfigure[$t = 2.0\,\mathsf{s}$]{\includegraphics[width=0.49\textwidth]%
  {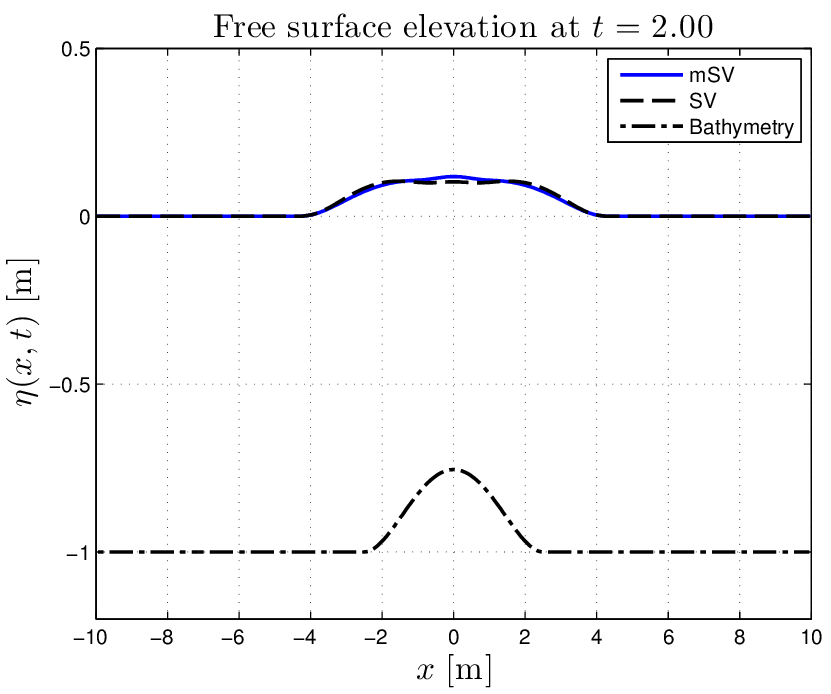}}
  \subfigure[$t = 5.0\,\mathsf{s}$]{\includegraphics[width=0.49\textwidth]%
  {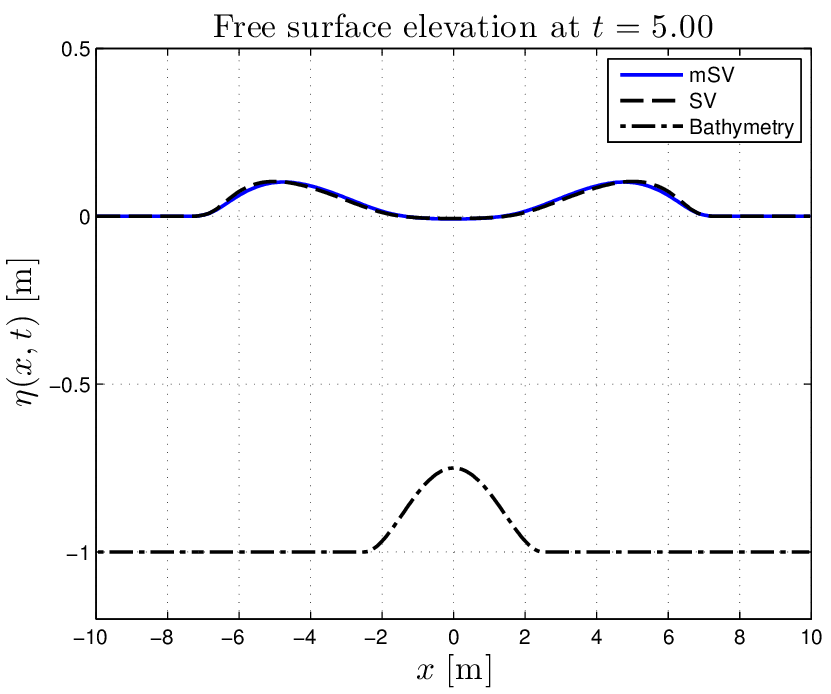}}
  \caption{\small\em Slow bottom uplift test-case ($\alpha_1 = 2$).}
  \label{fig:a2t200-500}
\end{figure}

\begin{figure}
  \centering
  \subfigure[$t = 0.5\,\mathsf{s}$]{\includegraphics[width=0.49\textwidth]%
  {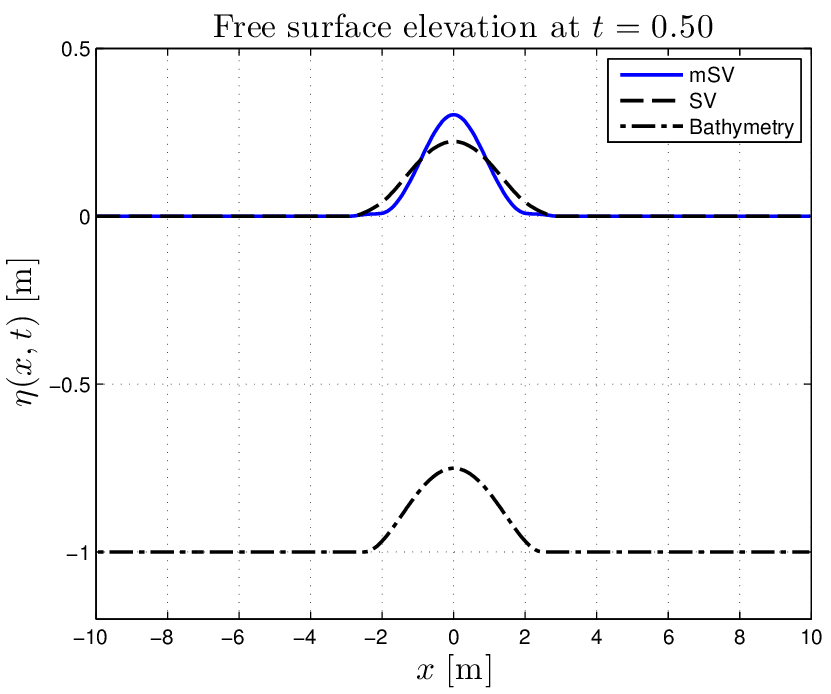}}
  \subfigure[$t = 0.9\,\mathsf{s}$]{\includegraphics[width=0.49\textwidth]%
  {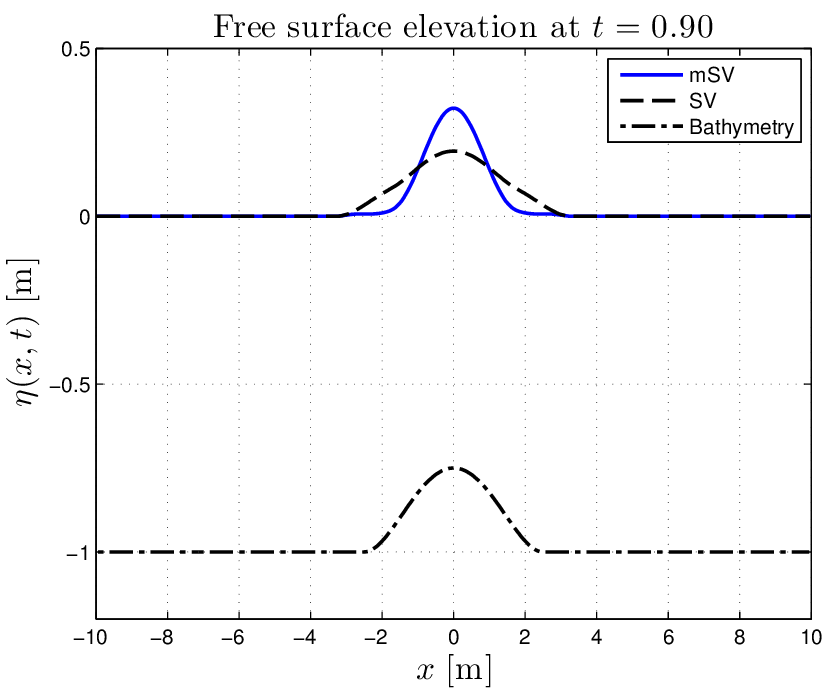}}
  \caption{\small\em Fast bottom uplift test-case ($\alpha_2 = 12$).}
  \label{fig:a12t050-090}
\end{figure}

\begin{figure}
  \centering
  \subfigure[$t = 1.0\,\mathsf{s}$]{\includegraphics[width=0.49\textwidth]%
  {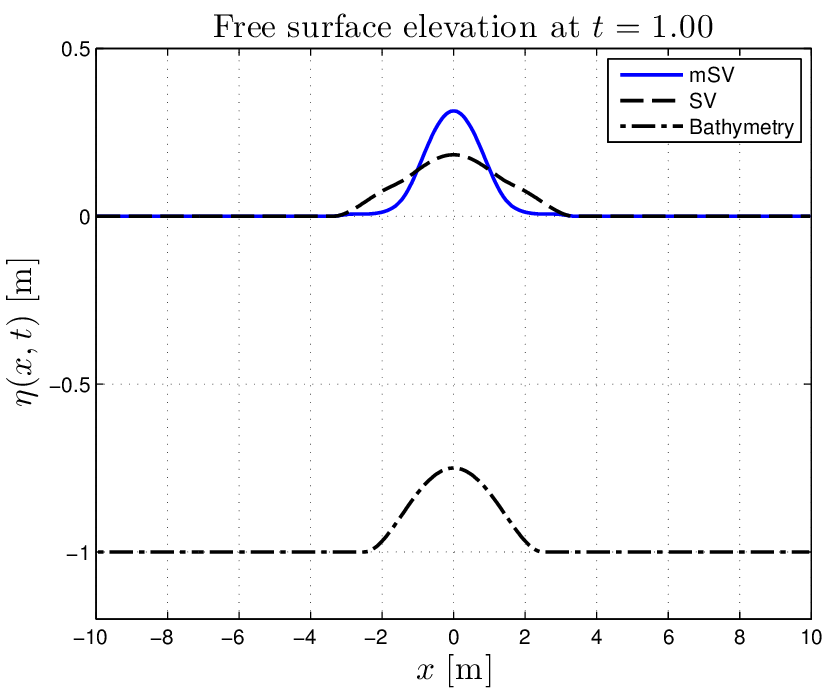}}
  \subfigure[$t = 1.5\,\mathsf{s}$]{\includegraphics[width=0.49\textwidth]%
  {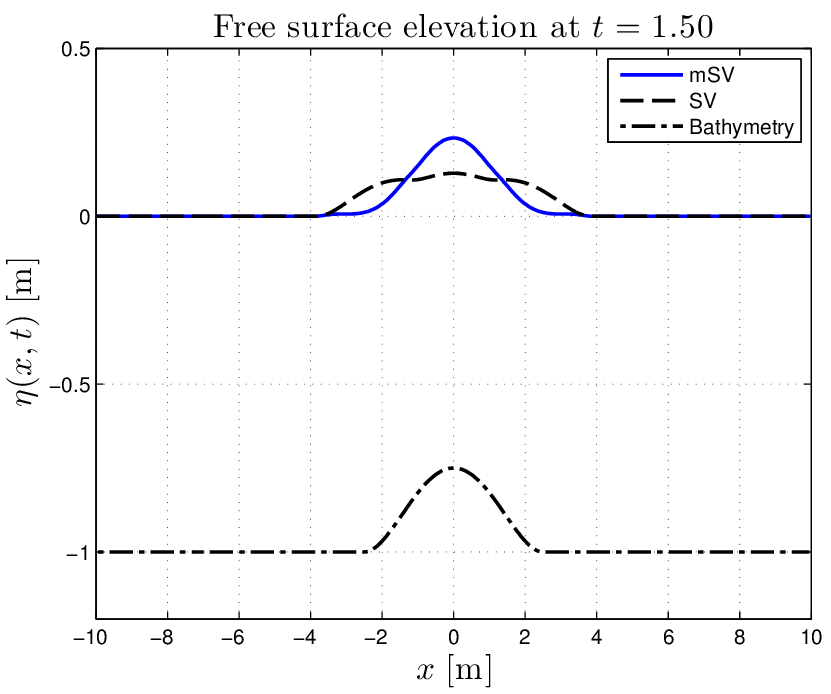}}
  \caption{\small\em Fast bottom uplift test-case ($\alpha_2 = 12$).}
  \label{fig:a12t100-150}
\end{figure}

\begin{figure}
  \centering
  \subfigure[$t = 2.0\,\mathsf{s}$]{\includegraphics[width=0.49\textwidth]%
  {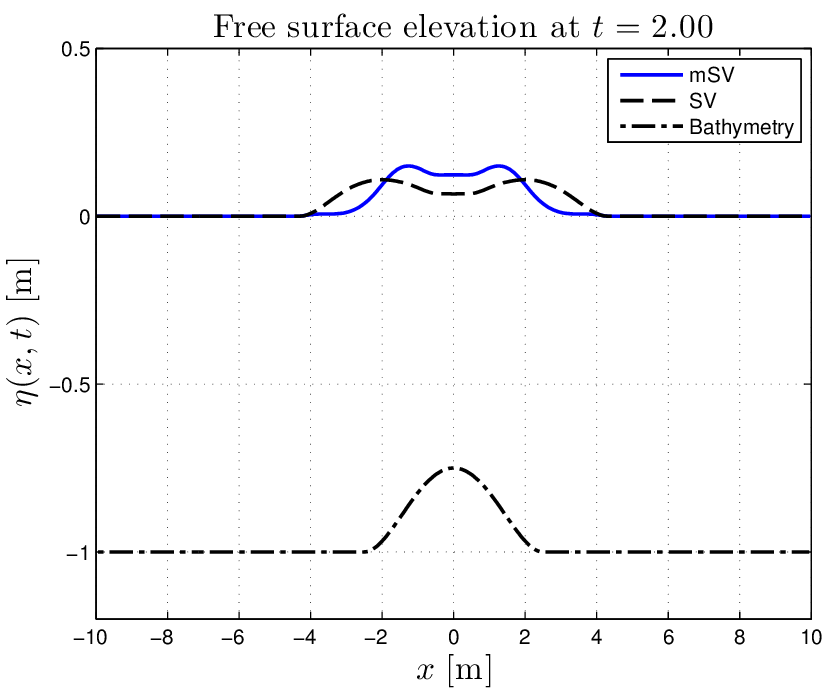}}
  \subfigure[$t = 2.5\,\mathsf{s}$]{\includegraphics[width=0.49\textwidth]%
  {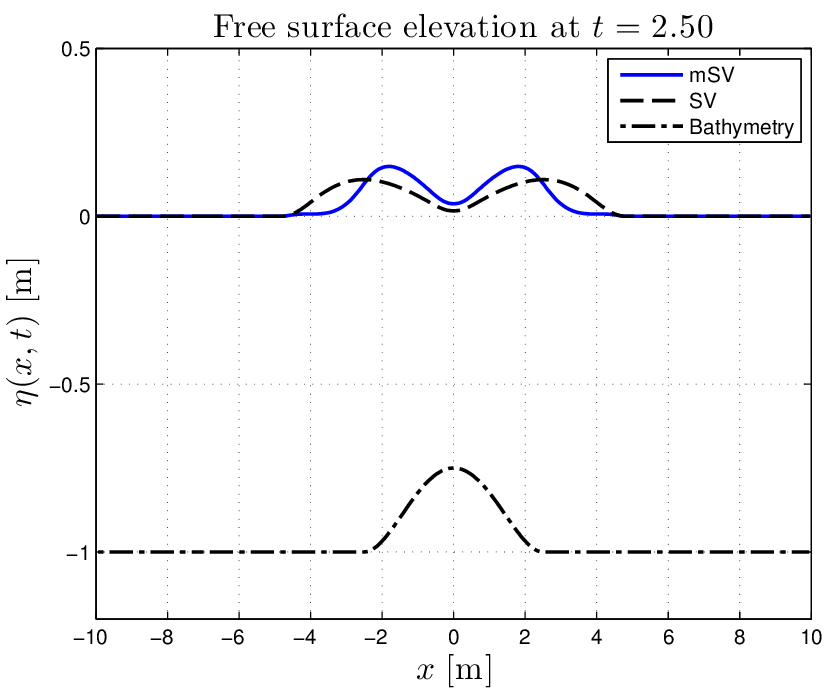}}
  \caption{\small\em Fast bottom uplift test-case ($\alpha_2 = 12$).}
  \label{fig:a12t200-250}
\end{figure}

\begin{figure}
  \centering
  \subfigure[$t = 3.0$ s]{\includegraphics[width=0.49\textwidth]%
  {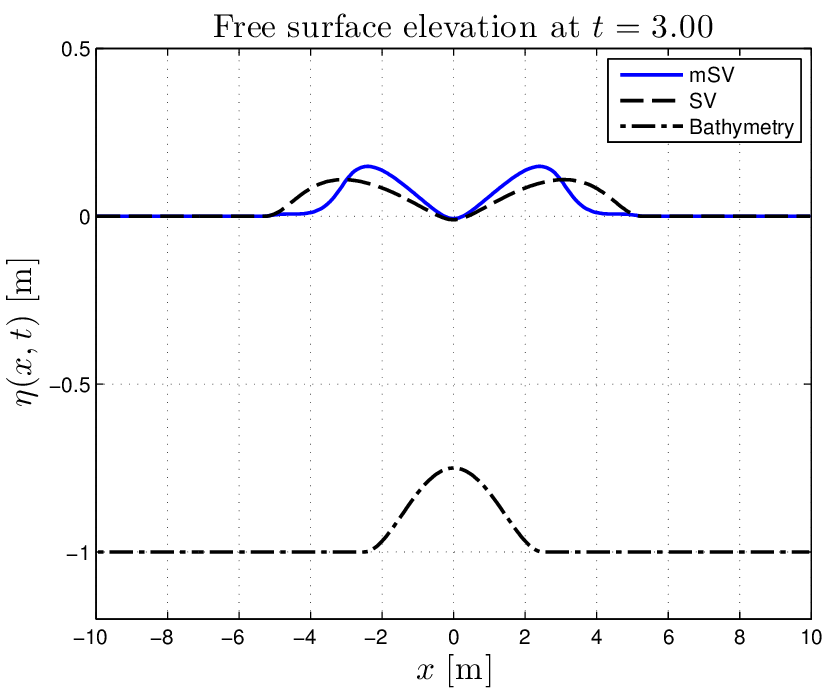}}
  \subfigure[$t = 5.0$ s]{\includegraphics[width=0.49\textwidth]%
  {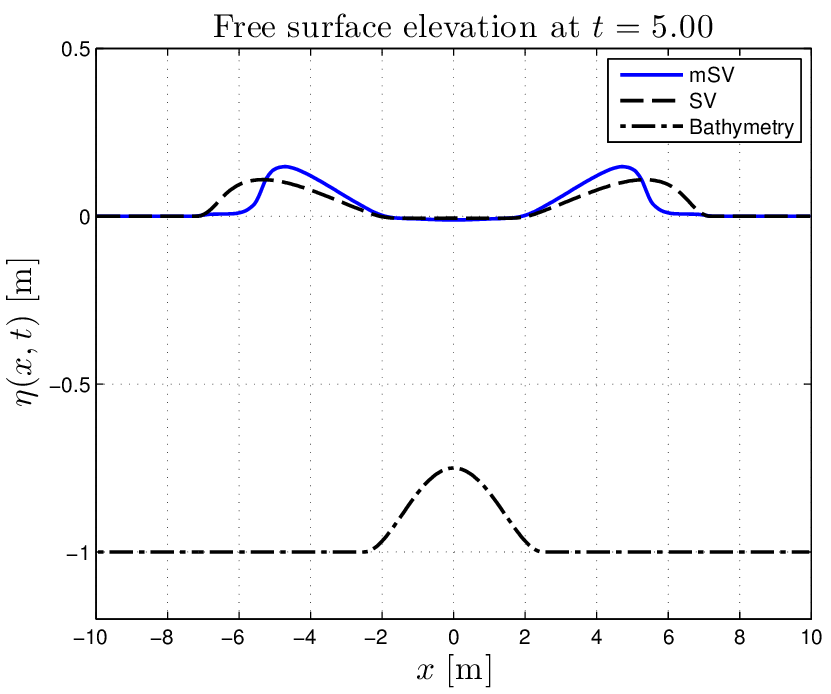}}
  \caption{\small\em Fast bottom uplift test-case ($\alpha_2 = 12$).}
  \label{fig:a12t300-500}
\end{figure}

Numerical results of the moving bottom test-case are shown in Figures~\ref{fig:a2t50-100}--\ref{fig:a12t300-500}. On all these images the blue solid line corresponds to the \acs{msv} equations, while the black dashed line refers to its classical counterpart. The dash-dotted line shows the bottom profile which evolves in time as well.

First, we present numerical results (see Figures~\ref{fig:a2t50-100}--\ref{fig:a2t200-500}) corresponding to a relatively slow uplift of a portion of the bottom ($\alpha_1 = 2.0$). There is a very good agreement between two computations. We note that the amplitude of bottom deformation $a/d = 0.25$ is strong which explains some small discrepancies in Figure~\ref{fig:a2t200-500}(a) between two models. This effect is rather due to the bottom shape than to its dynamic motion.

Then we test the same situation but the bottom uplift is fast with the inverse characteristic time $\alpha_2 = 12.0$. In this case the differences between two models are very flagrant. As it can be seen in Figure~\ref{fig:a12t100-150}, for example, the \acs{msv} equations give a wave with almost two times higher amplitude. Some differences in the wave shape persist even during the propagation (see Figure~\ref{fig:a12t300-500}). This test-case clearly shows another advantage of the modified Saint-Venant equations in better representation of the vertical velocity field.

\begin{figure}
  \centering
  \subfigure[\it Slow uplift ($\alpha_1$)]{\includegraphics[width=0.49\textwidth]%
  {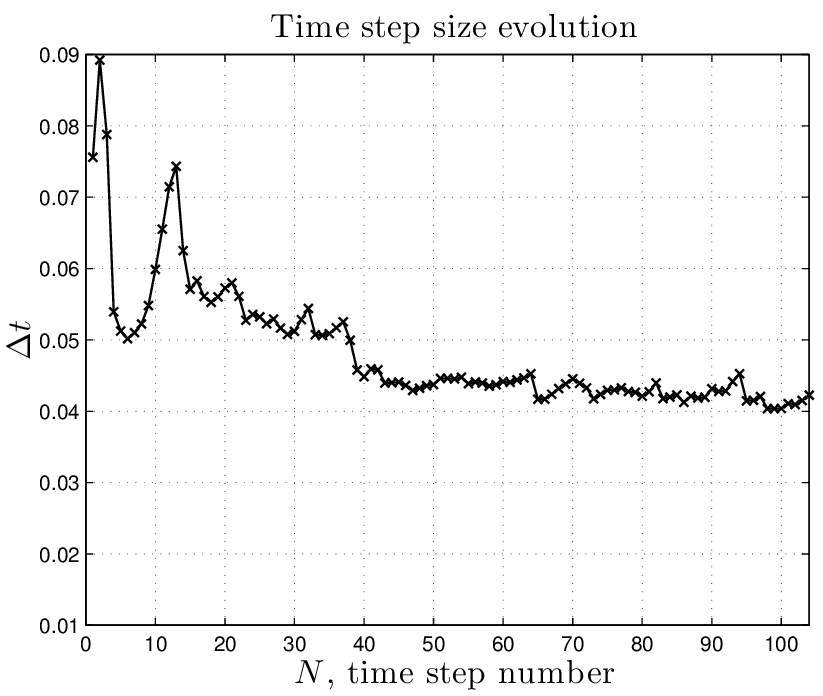}}
  \subfigure[\it Fast uplift ($\alpha_2$)]{\includegraphics[width=0.49\textwidth]%
  {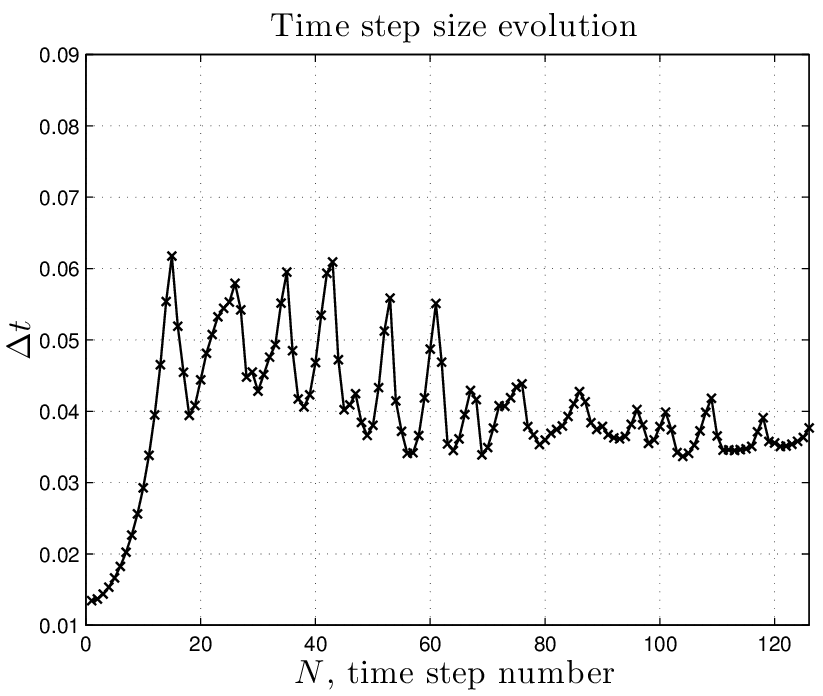}}
  \caption{\small\em Local time step size evolution during the simulation of a wave generation by moving bottom.}
  \label{fig:timegen}
\end{figure}

On Figure \ref{fig:timegen} we show the evolution of the local time step adapted while solving the \acs{msv} equations with moving bottom (up to $T = 5\,\mathsf{s}$). We can observe a behaviour very similar to the result presented above (see Figure~\ref{fig:timestep}) for the wave propagation test-case.

\subsection{Application to tsunami waves}

Tsunami waves continue to pose various difficult problems to scientists, engineers and local authorities. There is one question initially stemming from the Ph.D. thesis of C.~\textsc{Synolakis} \cite{Synolakis1986}. On page 85 of his manuscript, one can find a comparison between a theoretical (NSWE) and an experimental wavefront paths during a solitary wave runup onto a plane beach. In particular, his results show some discrepancy whose importance was not completely recognized until the wide availability of videos of the Tsunami Boxing Day 2004 \cite{Ammon2005, indiens2, Titov}. In the same line of thinking, we quote here a recent review by Synolakis and Bernard \cite{Syno2006} which contains a very interesting paragraph:
\begin{quote}
{\em ``In a video taken near the Grand Mosque in Aceh, one can infer that the wavefront first moved at speeds less than 8 km h$^{-1}$, then accelerated to 35 km h$^{-1}$. The same phenomenon is probably responsible for the mesmerization of victims during tsunami attacks, first noted in series of photographs of the 1946 Aleutian tsunami approaching Hilo, Hawaii, and noted again in countless photographs and videos from the 2004 mega-tsunami. The wavefront appears slow as it approaches the shoreline, leading to a sense of false security, it appears as if one can outrun it, but then the wavefront accelerates rapidly as the main disturbance arrives.''}
\end{quote}

Since our model is able to take into account the local bottom slope into the wave speed computation, we propose below a simple numerical setup which intends to shed some light on possible mechanisms of the reported above wave front propagation anomalies. Consider a one-dimensional domain $[-20,20]$ with wall boundary conditions. This domain is discretized into $N = 4000$ control volumes in order to resolve local bathymetry oscillations. The bottom has a uniform slope which is perturbed on the left side ($x < 0$) by fast oscillations which model the bottom ``steepness''
\begin{equation}\label{eq:dxs}
  d(x)\ =\ d_0\ -\ x\tan(\delta)\ +\ a\,[\,1\,-\operatorname{H}(x)\,]\sin(kx),
\end{equation}
where $\operatorname{H}(x)$ is the Heaviside function. The initial condition is a solitary wave moving rightwards as it was chosen in \cite{Synolakis1986}:
\begin{equation*}
  \frac{\eta_0(x)}{d(x_0)}\ =\ A\sech^2\!\left(\half\kappa(x-x_0)\right), \qquad u_0(x)\ =\ \frac{c_0\,\eta_0(x)}{d(x_0) + \eta_0(x)},
\end{equation*}
\begin{equation*}
  \kappa d(x_0)\ =\ \sqrt{\frac{3\,A}{1+A}}, \qquad \frac{c_0^{\,2}}{g\,d(x_0)}\ =\ 1\,+\,A.
\end{equation*}
This configuration aims to model a wave transition from steep to gentle bottoms. The values of various physical parameters are given in Table~\ref{tab:front}.

\begin{table}
  \centering
  \begin{tabular}{l|c}
    \hline\hline
    \textit{Parameter} & \textit{Value} \\
    \hline\hline
    Undisturbed water depth $d_0$:  & $1\,\mathsf{m}$ \\
    Gravity acceleration $g$:       & $1\,\mathsf{m\,s^{-2}}$ \\
    Bottom slope $\tan(\delta)$:    & $0.02$ \\
    Oscillation amplitude $a$:      & $0.1\,\mathsf{m}$ \\
    Oscillation wavenumber $k$:     & $20\,\mathsf{m^{-1}}$ \\
    Final simulation time $T$:      & $19\,\mathsf{s}$ \\
    Solitary wave amplitude $A$:    & $0.3\,\mathsf{m}$ \\
    Solitary wave initial position $x_0$:   & $-12.0\,\mathsf{m}$ \\
    \hline\hline
  \end{tabular}
  \bigskip
  \caption{\small\em Values of various physical parameters used for the wave propagation over a sloping bottom.}
  \label{tab:front}
\end{table}

\begin{figure}
  \centering
  \includegraphics[width=0.58\textwidth]{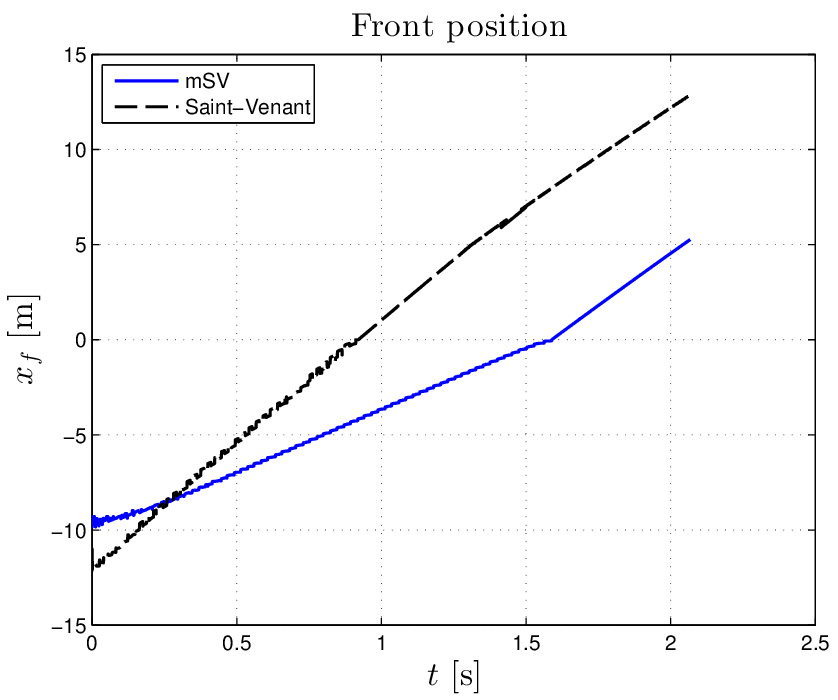}
  \caption{\small\em Wave front position computed with modified and classical Saint-Venant equations.}
  \label{fig:front}
\end{figure}

Then, the wave propagation and transformation over the sloping bottom \eqref{eq:dxs} was computed using the classical and modified Saint-Venant equations. The wave front position was measured along this simulation and the computation result is presented in Figure~\ref{fig:front}. The slope of these curves represents physically the wave front propagation speed. Recall also that the point $x = 0$ corresponds to the transition between steep and gentle regions of the sloping beach.

As one can expect, the classical model does not really `see' a region with bathymetry variations, except from tiny oscillations. An observer situated on the beach, looking at the upcoming wave modeled by the classical Saint-Venant equations, will not see any change in the wave celerity. More precisely, the slope of the black dashed curve in Figure~\ref{fig:front} is rather constant up to the graphical resolution. On the other hand, one can see a drastic change in the wave front propagation speed predicted by the \acs{msv} equations when the bottom variation disappear.

The scenario we present in this section is only a first attempt to shed some light on the reported anomalies in tsunami waves arrival time on the beaches. For instance, a comprehensive study of P.~\textsc{Wessel} \cite{Wessel2009} shows that the reported tsunami travel time is often exceeds slightly the values predicted by the classical shallow water theory (see, for example, Figures~5 and 6 in \cite{Wessel2009}). This fact supports indirectly our theory. Certainly this mechanism does not apply to laboratory experiments but it can be a good candidate to explain the wave front anomalies in natural environments. The mechanism we propose is only an element of explanation. Further investigations are needed to bring more validations to this approach.

We underline that the computational results rely on sound physical modeling without any {\em ad hoc} phenomenological terms in the governing equations. Only an accurate bathymetry description is required to take the full advantage of the \acs{msv} equations.

\section{Conclusions}\label{sec:concl}

In this study, we derived a novel non-hydrostatic non-dispersive model of shallow water type which takes into account large bathymetric variations. Previously, some attempt was already made in the literature to derive shallow water systems for arbitrary slopes and curvature \cite{Berger1998, Bouchut2003, Dressler1978, Keller2003}. However, our study contains a certain number of new elements with respect to the existing state of the art. Namely, our derivation procedure relies on a generalised Lagrangian principle of the water wave problem \cite{Clamond2009} which allows easily the derivations of approximations that cannot be obtained with more conventional asymptotic expansions. Indeed, we do not introduce explicitly any small parameter and our approximation is made through the choice of a suitable ansatz. Resulting governing equations have a simple form and physically sound structure. Another new element is the introduction of arbitrary bottom time variations. Finally, the non-hydrostatic character of obtained equations is fundamentally different from the well-known Boussinesq-type and \textcolor{red}{mild-slope} models. The reason of non-hydrostaticity of mSV equations lies in the pressure term and not in the frequency dispersion. \textcolor{red}{Of course, the proposed model has to be further tested and validated by making direct comparisons with state-of-the-art numerical wave models \cite{Grilli2008, Athanassoulis2015}.}

The proposed model is discretised with a finite volume scheme with adaptive time stepping to capture the underlying complex dynamics. The performance of this scheme is then illustrated on several test cases. Some implications to tsunami wave modelling are also suggested at the end of this study. For ocean modelling, the most interesting feature of the model is perhaps the prediction of a wave slow down due to the bottom slope.

Among various perspectives, we would like to underline the importance of a robust runup algorithm development using the current model. This research should shift forward the accuracy and our comprehension of a water wave runup onto complex shores \cite{Dutykh2011e, Dutykh2011c}.


\subsection*{Acknowledgments}
\addcontentsline{toc}{subsection}{Acknowledgments}

The authors acknowledge the support from CNRS under the PEPS InPhyNiTi project \texttt{FARA}. We would like to thank also Professors Valeriy \textsc{Liapidevskii}, Dimitrios \textsc{Mitsotakis} and Theodoros \textsc{Katsaounis} for interesting discussions on gravity driven currents and finite volume schemes. We also thank Professor Thierry \textsc{Gallou\"{e}t} for the suggestion to take into consideration steady solutions.


\addcontentsline{toc}{section}{References}
\bibliographystyle{abbrv}
\bibliography{biblio}

\end{document}